\documentclass{aa}
\usepackage[varg]{txfonts}
\usepackage{graphicx}
\usepackage{epstopdf}

\usepackage{natbib}
\usepackage{multirow}

\usepackage[squaren,Gray]{SIunits}
\usepackage{subfig}

\bibpunct{(}{)}{;}{a}{}{,}

\newcommand{\funit}{\,ph\,cm$^{-2}$\,s$^{-1}$}
\newcommand{\iunit}{\,ph\,cm$^{-2}$\,s$^{-1}$\,sr$^{-1}$}
\newcommand{\msol}{\,M$_{\odot}$}

\newcommand{\kev}{\,keV}
\newcommand{\mev}{\,MeV}

\newcommand{\gray}{$\gamma$-ray}

\newcommand{\nickel}{$^{56}$Ni}
\newcommand{\cobalt}{$^{56}$Co}
\newcommand{\alu}{$^{26}$Al}
\newcommand{\titan}{$^{44}$Ti}
\makeatletter

\newcommand{\Rmnum}[1]{\expandafter\@slowromancap\romannumeral #1@}
\makeatother
\begin{document}


\title{Monte Carlo modelling of the propagation and annihilation \\  
of nucleosynthesis positrons in the Galaxy} 

\author{A.~Alexis\inst{1,2}
  \and P.~Jean\inst{1,2}
     \and P.~Martin\inst{1,2}
     	\and K.~Ferri\`ere\inst{1,2}}

\offprints{A. Alexis, \email{aalexis@irap.omp.eu}}

\institute{Universit\'e de Toulouse; UPS-OMP; IRAP; Toulouse, France
\and CNRS ; IRAP ; 9, Avenue du Colonel Roche, BP 44346, F-31028 Toulouse cedex 4, France }

\date{Received 30 July 2013 / Accepted 23 February 2014}

\abstract 
{}
 {We want to estimate whether the positrons produced by the $\beta^{+}$-decay of \alu,\titan\ and \nickel\ synthesised in massive stars and supernovae are sufficient to explain the 511 \kev\ annihilation emission observed in our Galaxy. Such a possibility has often been put forward in the past. In a previous study, we showed that nucleosynthesis positrons cannot explain the full annihilation emission. Here, we extend this work using an improved propagation model.}
  {We developed a Monte Carlo Galactic propagation code for $\sim$MeV positrons in which the Galactic interstellar medium, the Galactic magnetic field and the propagation are finely described. This code allows us to simulate the spatial distribution of the 511 \kev\ annihilation emission. We test several Galactic magnetic fields models and several positron escape fractions from type-Ia supernova for \nickel\ positrons to account for the large uncertainties in these two parameters. We consider the collisional/ballistic transport mode and then compare the simulated 511\kev\ intensity spatial distributions to the INTEGRAL/SPI data.}
  {Whatever the Galactic magnetic field configuration and the escape fraction chosen for \nickel\ positrons, the 511 \kev\ intensity distributions are very similar. The main reason is that $\sim$MeV positrons do not propagate very far away from their birth sites in our model. The direct comparison to the data does not allow us to constrain the Galactic magnetic field configuration and the escape fraction for \nickel\ positrons. In any case, nucleosynthesis positrons produced in steady state cannot explain the full annihilation emission. The comparison to the data shows that: (a)  the annihilation emission from the Galactic disk can be accounted for; (b) the strongly peaked annihilation emission from the inner Galactic bulge can be explained by positrons annihilating in the central molecular zone but this seems to require more positron sources than the population of massive stars and type Ia supernovae usually assumed for this region;  (c) the more extended emission from the Galactic bulge cannot be explained. We show that a delayed 511 \kev\ emission from a transient source, such as a starburst episode or a recent activity of Sgr A*, occurring between 0.3 and 10 Myr ago and producing between 10$^{57}$ and 10$^{60}$ sub-MeV positrons could explain this extended component, and potentially contribute to the inner bulge signal. }
{}

\keywords{Astroparticle physics -- Gamma rays: ISM -- Nuclear reactions, nucleosynthesis, abundances}
\maketitle



\section{Introduction}
\label{intro}

The 511 keV line emission from our Galaxy is unambiguously produced by low energy positrons that annihilate with electrons, but the exact origin of these positrons remains unclear. The spatial distribution of the annihilation emission was measured by several generations of gamma-ray instruments and most recently by INTEGRAL/SPI\footnote{For more information on SPI, see \citet{Vedrenne03}.} \citep{Knodlseder05,Weidenspointner08b,Bouchet10}. It comprises faint emission from the inner part of the Galactic disk (GD), and a strong diffuse emission from the Galactic bulge \citep[GB, which can be modelled with a narrow and a wide spheroidal gaussian distribution with projected FWHM of $\sim$3$\degree$ and $\sim$11$\degree$ respectively, see][]{Weidenspointner08b}. This emission is very particular, with an inferred bulge-to-disk luminosity ratio ranging from 2 to 6. None of the known Galactic astrophysical object or interstellar matter distributions resembles the annihilation emission distribution.

In order to explain this bulge-to-disk luminosity ratio, several authors suggested that positrons produced by supernovae in the disk could propagate far enough to annihilate in the bulge \citep{Prantzos06,Higdon09}. Other authors proposed that mini-starbursts or the supermassive black hole Sgr A* produced a large amount of positrons 10$^6$ or 10$^7$ years ago, which filled the GB and are annihilating now \citep{Parizot05,Totani06,Cheng06,Cheng07}.

A major issue in such studies is that the propagation of positrons in the
interstellar medium (ISM) is not well understood \citep{Jean06}. 
In a previous detailed analysis, \citet{Higdon09} used an inhomogeneous 
diffusion model, including collisional and collisionless transport of positrons,
in a finely-structured ISM and found that nucleosynthesis positrons could account 
for all the observables of the Galactic annihilation emission.
However, this model raised some criticism in the community 
\mbox{\citep[e.g.,][]{Prantzos11,Martin12}.} 
In the present paper, we would like to propose a different approach 
based on the theoretical investigation of \citet{Jean09} on propagation mechanisms for positrons in the ISM.
These authors showed that at low energy (E $\lesssim$ 10 MeV), positrons do not interact with magnetohydrodynamic waves and propagate in the collisional mode by undergoing gyro-motion around magnetic field lines and collisions with gas particles. In such conditions, a correct treatment of positron propagation requires using representative models of Galactic magnetic field (GMF) and Galactic gas distributions. In a previous study, we performed a simulation of the propagation of positrons emitted in the decay of radioactive nuclei produced by massive stars and supernovae (\alu, \titan\ and \nickel) using a modified version of the GALPROP cosmic-ray propagation code \citep{Martin12}. This code treats the transport of positrons as a diffusive process and uses 2D analytical distributions of the large-scale average gas density. In this framework, we showed that it is hard to explain the morphology of the annihilation emission from radioactivity positrons, which led us to the conclusions that either an additional component is needed to explain the bulge emission, or a finer modelling is required. We explored the latter option in the present work.

In this paper, we investigate the fate of positrons produced by stellar nucleosynthesis in our Galaxy with a Monte Carlo code that takes into account the transport of positrons in the collisional mode. This code simulates the injection, propagation, energy loss and annihilation of positrons taking into account spatial distributions for sources, interstellar gas and magnetic field. The results of the simulations allow us to derive sky maps, light curves and spectra of the annihilation emission as functions of the sources of positrons.
Section \ref{model} describes the Monte Carlo method and the various model components. In Section \ref{simulations}, we present and discuss the results of the simulations. The simulated sky maps of the annihilation emission at 511 keV are compared to the INTEGRAL/SPI data. We show that type Ia supernovae (SNe Ia) cannot be the main source of positrons and that additional sources are needed to explain the measured disk and bulge emissions. The latter can be explained by a brief injection of a large amount of positrons in the central region of our Galaxy (e.g. in the central molecular zone) that occurred several Myr ago. In Section \ref{conclusion}, we summarize our study and conclusions.


\section{The Monte Carlo Galactic propagation model}
\label{model}

The 511\kev\ annihilation emission mainly depends on four inputs: the properties of the ISM, the configuration of
the GMF, the positron source spatial and spectral distribution, and the positron propagation physics.
In the following, we introduce the models and assumptions used for each of these inputs. We then summarise the development of the Monte Carlo simulations and explain how 511 \kev\ intensity sky maps were generated.

\subsection{Modelling the interstellar medium}
\label{ism}

We consider a static model of the ISM, which does not include any dynamic
phenomena such as Galactic winds, chimneys, etc...
We divided the Galaxy into three regions: the Galactic disk (GD, $R$ $\geq$~1.5 kpc,  with $R$
the Galactocentric radius), the Galactic Bulge (GB, 0.01 kpc < $R$ < 1.5~kpc), and the Sgr A* region \mbox{($R$ $\leq$ 10 pc).}
We then used the spatial distribution of the interstellar gas given by  \citet{Ferriere98} and \citet{Ferriere07} for the GD and the GB, respectively.
In these models, the ISM, composed of 90\% hydrogen and 10\% helium, is described by five gaseous phases: the molecular medium (MM), the cold neutral medium (CNM), the warm neutral medium (WNM), 
the warm ionized medium (WIM) and the hot ionized medium (HIM). 
These models give the space-averaged density $\langle n_{i}\rangle$ of each ISM phase $i$.
In the GB model, the neutral (molecular and atomic) gases are confined to two structures:
the so-called Central Molecular Zone (CMZ) and a holed tilted disk. 
The CMZ is a 500~pc $\times$ 200~pc ellipse with a FWHM thickness of 30~pc,
while the holed tilted disk is a 3.2~kpc $\times$ 1~kpc ellipse with 2.3 times 
the FWHM thickness of the CMZ and with a hole in the middle to leave room for the CMZ.
Because the CNM and the WNM in the GB cannot be separated observationally,
the GB model only gives the space-averaged density
of the total (CNM + WNM) atomic gas.
Let us denote by $f_{\rm CNM}$ and $f_{\rm WNM}$ the respective fractions
of atomic hydrogen (HI) space-averaged density in the form of CNM and WNM.
Due to the high thermal pressure and ionization rate in the
GB \citep[e.g.,][]{Morris96}, we expect most of the atomic gas 
to be in the form of CNM. Indeed, thermal pressure is almost certainly 
above the critical pressure for the existence of warm atomic gas 
under thermal equilibrium conditions, and while departures from
thermal equilibrium may allow for the presence of some warm gas,
part of it will surely be ionized by the high ionization rate.
Here, we adopted the conservative estimates $f_{\rm CNM}=0.7$ 
and $f_{\rm WNM}=0.3$. Finally, we derived the true density, $n_i$, of
each phase from its space-averaged density, $\langle n_i \rangle$,
using the method described in Jean et al. (2006): we multiplied its true density near the Sun $n_{i,\odot}$
by a common ``compression factor'', $f_\mathrm{c}$, adjusted to ensure that 
$\sum_{{i}} \phi_{i} = 1$ where $\phi_{i}=\langle n_{i}\rangle/n_{i}$ 
is the volume filling factor of phase $i$.
 For the ionization fraction and the temperature of each phase,
 we took the mean values given in Table 1 of \citet{Jean09}. 
  
To obtain a complete model of the Galaxy, we also modelled the interstellar gas within $\sim$10 pc of  Sgr A$^{*}$ following the recent prescription
by \citet{Ferriere12}. The ISM components of the Sgr A$^{*}$ region are geometrically identified and modelled with their thermodynamic
parameters \citep[see Table 2 of][]{Ferriere12}. In brief, this region can be seen as an ionized radio halo (IRH) enclosing a warm ionized central
cavity, the Sgr A East supernova remnant (SNR) and a multitude of molecular structures.

We simulated positron propagation in a finely-structured ISM. 
During propagation, a positron successively goes through clearly identified phases, and a choice had to be made about the transition from one phase to another.
The Galaxy contains some regions with a well-defined structure of phases, called photodissociation regions \citep{Tielens85}, where atomic/molecular clouds are illuminated by strong ultraviolet radiation fields and have their outer layers largely ionized. These regions are particularly found in the Galactic nucleus and in the CMZ \mbox{\citep[see e.g.,][]{Wolfire90}}. Yet, observations \citep[e.g.,][]{Heiles03} and numerical studies \citep[e.g.,][]{DeAvillez04,DeAvillez05} also show an ISM with a phase continuum, where the ISM is mixed down to relatively small scales, and not an ISM with a clear-cut separation between phases \citep[see][for a review on this topic]{VazquezSemadeni09}. Therefore, in this study, we considered two simplified extreme models for the Galactic ISM.

In the first model, which we called the random ISM model, each time a positron leaves an ISM phase, the next phase it enters is selected based on its filling factor.
Thus, each phase with non-zero filling factor has a finite probability of being entered by the positron. More details are given in Sect. \ref{physics}.

In the second model, which we called the structured ISM model, the different ISM phases are related to each other everywhere in the Galaxy. 
The basic structure that we considered is a spherical structure with increasing temperature and ionization fraction from the centre to the edge: a MM core, surrounded by a layer of CNM, itself surrounded by a layer of WNM, itself surrounded by a layer of WIM, with an outer envelope of HIM. In this model, a MM region cannot be found directly next to a HIM region. The volumes, and hence the radii, of the different phases of this structure are determined by their respective filling factors. 
Each time a positron escapes such a spherical structure, another structure is generated. The calculation of the exact dimensions of the spherical structure is presented 
in Sect. \ref{physics}.

These two representations are limiting cases of the layout and ordering of the ISM, and we will discuss their respective impact on the results.

\subsection{Modelling the Galactic magnetic field}
\label{gmf}

The structure of the GMF is often described
with two components: the regular GMF and the turbulent GMF. These two components are probed with
measurements of the total and polarized synchrotron emission and Faraday
rotation of pulsars and extragalactic sources. 
Recent studies \citep{Sun08,Sun10,Jansson12a} 
tend to show that the regular GMF could be made up of a disk field and a halo
field component. 

We modelled the disk regular component using the model
of \citet{Jaffe10}, which is  a parametric two-dimensional coherent spiral arm 
magnetic field model which provides predictions for observables such as
synchrotron intensities and Faraday rotation measures. 
In order to obtain a complete three-dimensional model of the 
regular GMF, we assumed that the spiral field strength decreases exponentially above
and below the Galactic plane with a scale height of 1 kpc \citep[see e.g.][]{Prouza03,Sun08}.

The configuration of the halo regular field is even more uncertain
than that of the disk regular field. This
halo component has been suggested to be a poloidal field with
a dipole shape \citep[see][for a review]{Han04}. Based on this dipole morphology, \citet{Prantzos06} 
argued that the positrons produced in the disk could
be transported into the bulge, thereby
explaining the atypical 511 keV emission distribution.
However, the recent analysis by \cite{Jansson12a} found support for the presence of an X-shape magnetic field
in the Galactic halo, as could be expected from observations
of X-shape fields in external spiral galaxies seen edge-on \citep[see e.g.][]{Krause09}. Due to the uncertainties in our knowledge of
the halo GMF, we tested three configurations of the halo
regular field: no halo field, the dipole field as described in \citet{Prouza03} and the X-shape field as described in \citet{Jansson12a}.
The three configurations will be respectively denoted by N, D and X in the Tables.

The status of the GMF in the Sgr A* region is rather uncertain and has never been thoroughly reviewed. In this work, we assumed that 
the magnetic field in all the diffuse and ionized regions near Sgr A* is perpendicular to the Galactic plane and has a strength of 
0.1 mG  \citep[see][for a description of the Sgr A* region]{Ferriere12}. 
We then assumed that the magnetic field in the dense and neutral regions is oriented along the long dimension of
the local clouds and has a strength of 1 mG \citep[][and references therein]{Ferriere09}.

In addition to the regular component, we modelled the turbulent GMF using the plane
wave approximation method described by \citet{Giacalone94}. 
We assumed that magnetic field fluctuations follow a Kolmogorov spectrum and have a maximum turbulent scale
$\lambda{}_{\rm max}$ in the range 10-100 pc in the hot and warm ISM phases and 1-10 pc
 in the cold neutral and molecular phases of the ISM. 
In previous studies on the positron propagation~\citep{Prantzos06,Jean09}, the ratio $\rm \delta{}B/B_{0}$ was assumed to 
be constant throughout the Galaxy. Here, we allowed this ratio to vary in space in the GD 
with $\rm \delta{}B/B_{0}$  increasing smoothly from 1 in interarm regions to 2 
along the arm ridges \citep[see][]{Jaffe10}. We set this ratio to 1 in the GB.

\subsection{Modelling the positron sources}
\label{sources}

The most promising source of Galactic positrons is the $\beta^{+}$-decay of unstable nuclei synthesised in massive stars or supernovae.
The reasons for this are: (a) some radio-isotopes emitting positrons, such as \alu\ and \titan, have been observed within the Galaxy
via the gamma-ray or X-ray lines that accompany their decay, (b) the observed or theoretical radio-isotope yields can supply positrons so as to feed the 511 \kev\ luminosity derived from INTEGRAL observations, and (c) the positrons  from radioactivity are released in the ISM with energies on average
lower than 1\mev, which is in agreement with the constraints obtained by \citet{Beacom06} and \citet{Sizun06}.
In the following, we make a short summary of Sect. 4 of \citet{Martin12}, who present all of the properties of each source studied here, and we highlight
the slight differences with their work. We assumed a steady-state Galactic production rate of all the following radio-isotopes.

The radio-isotope \alu\ decays with a lifetime of $\sim$1 Myr, emitting a gamma-photon at 1809 \kev\ and a positron 82\% of the time. 
Its spatial distribution is strongly correlated with the free-free emission from HII regions surrounding massive stars \citep{Knodlseder99}, confirming
that its nucleosynthesis is linked to massive stars. We therefore used the free-electron spatial distribution (NE2001~model) of \citet{Cordes03} for the distribution of  \alu. 
More specifically, we adopted the thin disk and spiral arm components of the NE2001 model for the disk massive stars. In the following, 
we call this component the star-forming disk (SFD) component. We also took into account the Galactic center (GC) component from the NE2001 model, 
since roughly 10\% of the massive stars could be formed in the inner stellar bulge ($R$ < 0.2 kpc), following the argument given by \citet{Higdon09}. 
This component is quite similar to the CMZ defined in \citet{Ferriere07}, so in the following we call it the CMZ component and we assign it 10$\%$ of the \alu\ positrons. 
Due to the very long decay lifetime of \alu, positrons are very likely injected into the ISM  with their original $\beta^{+}$-spectrum with a mean energy of $\simeq$0.45~\mev.
The steady-state Galactic production rate of \alu\ positrons can be derived from the present-day mass equilibrium of \alu\ in the Galaxy.
Here, we took the value of $\simeq$(2.8$\pm$0.8)\msol\ \citep{Diehl06} but we also took note of the estimate of \mbox{1.7-2.0$\pm$0.2\msol} derived by \citet{Martin09}. 
With a $\beta^{+}$-decay branching ratio of 82 \%, we obtain a positron production rate $\simeq$(3.2$\pm0.9$)$\times10^{42}~\mathrm{e}^{+}/\mathrm{s}$.

The radio-isotope \titan\ decays with a lifetime of $\simeq$85 yr into $^{44}$Sc, which in turn decays very quickly into $^{44}$Ca,
 emitting a positron 94\% of the time. \titan\ is mainly synthesised during core-collapse supernova explosions (ccSNe) of massive stars.
 Thus, we used the same spatial distribution (SFD+CMZ) as for \alu\ for the spatial distribution of \titan. \titan\ positrons have to travel across stellar 
 ejecta before entering the ISM, but because of the intermediate decay lifetime of the radio-isotope, we assumed that \titan\ positrons are also released into the ISM with
their original $\beta^{+}$-spectrum with a mean energy of $\simeq$0.6 \mev. Based on the production rate of  {$^{56}$Fe} and the measured solar 
({$^{44}$Ca}~/{$^{56}$Fe})$_{\odot}$ ratio \citep[see][]{Prantzos11}, the positron production rate from \titan\ is $\simeq$3$\times 10^{42}~ 
 \mathrm{e}^{+}/\mathrm{s}$. We assumed an uncertainty range of $\pm$50$\%$ on this positron injection rate to reflect the uncertainties on the \titan\ production rate.
 
The radio-isotope \nickel\ decays with a lifetime of $\simeq$9 days into $^{56}$Co, which in turn decays with a lifetime of $\simeq$111 days into $^{56}$Fe, 
emitting a positron 19\% of the time. The \nickel\ is synthesised during ccSNe and thermonuclear supernova explosions (SNe Ia), but SNe Ia are by far the 
dominant source of positrons due to their higher iron yield per event and their much higher positron escape fraction from the ejecta \citep{Martin12}. 
\nickel\ therefore follows the time-averaged spatial distribution of SNe Ia in the Galaxy. \citet{Sullivan06} showed that the spatial distribution of SNe Ia is 
a combination of the young stellar populations and the stellar mass distributions. We thus assumed that a distribution of old/delayed SNe Ia follows an 
exponential disk (ED) with a central hole plus an ellipsoidal bulge (EB), both components tracing the stellar mass \citep[see Sect 6.1 of][]{Martin12}. 
Then, a population of early/prompt SNe Ia is associated with the SFD. We do not consider early/prompt SNe Ia occurring in the CMZ because the uncertainties 
on the SNe Ia rate in this region are large \citep[see for instance][]{Schanne07}. We will however discuss that point in Sect. \ref{comparison2data}.

\nickel\ and \cobalt\ have very short lifetimes and their positrons are injected directly into the ejecta of the supernova, very likely experiencing strong energy losses before reaching the ISM. Therefore, the $\beta^{+}$-~spectrum of \cobalt\ positrons is altered in comparison with the original $\beta^{+}$-spectrum. Using the method of \citet[Sect.~5]{Martin10}, we computed some altered $\beta^{+}$-spectra for three escape fractions from the ejecta: 0.5\%, 5\% and 10\%. We chose these three values 
because they lie in the range of the estimations of several studies \citep[see][for instance]{Chan93,Milne99}. 
The calculated altered $\beta^{+}$-spectra have a mean energy of 105, 175 and 205 \kev\ for the escape fraction of 0.5\%, 5\% and 10\%, respectively.
These values are very different from the mean energy of $\sim$0.6 \mev\ of the unaltered $\beta^{+}$-spectrum.
Using the same computation as \citet[Eq. 1]{Martin12} for the SN Ia occurrence rate, we derived a positron injection rate of
4.45, 4.17 and 6.0~$\times~10^{42}~\mathrm{e}^{+}/\mathrm{s}$ for the ED, the EB and the SFD component, respectively.
These values are given for a typical \nickel\ yield of 0.6 \msol\ per event, a $\beta^{+}$-decay branching ratio of 19\% and a positron escape fraction of 5\% from the 
stellar ejecta\footnote{ The \nickel\ positron production rate for a SN Ia escape fraction of 0.5$\%$ and 10$\%$ can be derived by multiplying the values cited in the text by a factor 0.1 and 2, respectively.}.


\subsection{Modelling the propagation physics}
\label{physics}

After being released in the ISM by their parent radio-isotope, positrons propagate within the Galaxy, slowing down until they annihilate directly
with an electron or via Positronium (Ps) formation. The Ps is the bound-state of a positron with an electron, which is formed 25\% of the time 
in the para-Ps state and 75\% of the time in the ortho-Ps state. The ortho-Ps decays in 140 ns into 3 photons of
energies totalling 1022 \kev\ and the para-Ps decays in 0.125 ns into 2 photons of 511 keV contributing to the 511 keV \gray\ which is
also produced by the direct annihilation of a positron with an electron \citep{Guessoum91,Guessoum05}.

A positron can propagate in the Galaxy under two different regimes: collisional or collisionless \citep{Jean09}.
In the collisional regime, the positron has a ballistic motion; it propagates spiralling along the Galactic magnetic field lines undergoing pitch
angle scattering due to collisions with gas particles. In the collisionless regime, the positron scatters off magneto-hydrodynamic 
waves associated with interstellar turbulence. \citet{Jean09} showed that positrons could only interact with the Alfv\'en wave turbulent cascade in the
ionized phases of the ISM, but the anisotropy of magnetic perturbations likely makes this transport mode inefficient \citep[see also][]{Yan02}.
In this study, we only considered the collisional transport mode. We took into account continuous energy-loss processes (Coulomb collisions, 
inverse Compton scattering, synchrotron), binary interactions with atoms and molecules (ionization and excitation) and
pitch angle scattering as described in Sect. 3 of \mbox{\citet{Jean09}}. 

A positron travels through the different phases of the ISM. In the random ISM model (as defined in Sect. \ref{ism}),
we assumed that the positron leaves a given phase when the distance travelled inside this phase is greater than a certain distance $d$ 
which is derived randomly from the probability density function of the distances that a particle can cross through a sphere of diameter $d_{i}$ in a straight line: 
\begin{equation}
\label{eq_dist}
d = \sqrt{\lambda} \times~d_{i}~~~~,
\end{equation}
where $\lambda$ is a random number uniformly distributed between 0 and 1, and $d_{i}$ is selected randomly in the typical size ranges 
of the considered ISM phase \citep[see Table 4 of][]{Jean06}.The new ISM phase $i$ is chosen randomly according to the probability 
\begin{equation}
\label{eq_phase1}
P_{i} = \frac{N_{i}~\sigma_{i}}{\sum_{j} ~N_{j}~\sigma_{j}} = \frac{\phi_{i}/d_{i}}{\sum_{j} ~\phi_{j}/d_{j}}~,
\end{equation}
where $\sigma_{i}$ is the cross section of the spherical region of phase $i$ ($\sigma_{i}$ = $\pi d_{i}^{2}/4$) and $N_{i}$
is the number density of spherical regions of phase $i$ ($N_{i} \propto \phi_{i} / V_{i}$, with $V_{i} = \pi d_{i}^{3}/6$ the volume of phase $i$). 

In the structured ISM model, the positron is injected at the surface of a new spherical structure. The radius of this structure, $r_{\rm sphere}$, is selected randomly between 50 and 100 pc, which roughly corresponds to the observed radii of evolved supernova remnants or the maximum sizes of the HIM \citep[see Table 4 of][]{Jean06}. In the CMZ, the radius $r_{\rm sphere}$ is selected randomly between 15 and 30 pc. With this range of $r_{\rm sphere}$ and a molecular gas filling factor \mbox{$\phi_{\rm MM}\simeq10-12\%$}, we find that
the innermost molecular region has a radius $\simeq$7--15 pc, consistent with the observed sizes of molecular clouds in the CMZ \citep[see, e.g.,][]{Oka98}.

At a given Galactic location, each phase filling factor $\phi_i$ is supposed to be known. In accordance with these filling factors, we fill the structure, from the outer surface to the centre, with ISM phases of decreasing temperature and ionization fraction. The radius of shell $s$ is thus given by
\begin{equation}
\label{eq_radius}
r_s = 
\left\lbrace
\begin{array}{ll}
r_{\rm sphere} \ ,& \quad s=4 \\
\left(r_{s+1}^{3}-r_{\rm sphere}^{3}\times\phi_{s+1}\right)^{\frac{1}{3}} \ ,& \quad s=0,1,2,3
\end{array}
\right.
,
\end{equation}
where  $s$ =0,1,2,3,4 refers to the MM, CNM, WNM, WIM, HIM, respectively. The ISM phases are thus fixed. This model locally reproduces the filling factors and specific transitions between the ISM phases. The positron is free to travel inside this onion skin structure until it escapes or annihilates. When the positron escapes, a new spherical structure is generated,
tangent to the previous spherical structure, with the new local filling factors of the different ISM phases.

\subsection{Summary of a Monte Carlo simulation}
\label{summary}

In the Monte Carlo code, the positron is first injected randomly in the Galaxy, at a certain location
depending on the initial spatial distribution of its radio-isotope, with a certain energy selected randomly from the original or altered $\beta^{+}$-spectrum of the radio-isotope 
(see Sect. \ref{sources}). 
We assume that the positron is released in the HIM with the direction of its initial velocity chosen randomly according to an isotropic velocity distribution.
Then, the positron propagates in the collisional regime following the turbulent and regular GMF lines (see Sect. \ref{gmf}), experiencing continuous energy losses,
pitch angle scattering and potentially binary interactions in the ISM phase in which it is travelling \citep[see Appendix B of][to have a complete overview of the
Monte Carlo algorithm]{Jean09}. During its lifetime, the positron changes ISM phase as described in Sect. \ref{physics}.
We emphasize that the collisional transport of the positron is truly 
inhomogeneous given that each ISM phase has its own regular magnetic field 
as a function of the current Galactic location (see Sect. \ref{ism}) 
and its own magnetic turbulence properties (see Sect. \ref{gmf}).

The tracking of a positron stops when it annihilates in-flight or when its energy drops below a threshold energy that we set to 100 eV,
below which the distance travelled by a positron is small compared to the size of any phase, except in the HIM
where we nevertheless keep on modelling the transport of the thermalized positrons. Due to the very low density of the HIM, thermal positrons
are very likely to escape the HIM and then to annihilate in a surrounding denser medium. 
Once the positron has annihilated, we store its final
position, propagation time, energy and ISM phase. The simulation can also stop when the positron escapes the Galaxy, i.e., when the positron goes higher than 5 kpc on either
side of the Galactic plane or beyond 20 kpc in Galactocentric radius.

By simulating a great number of positrons (N$_{0}$=10$^{5}$), we can estimate the steady state intensity spatial distribution at 511 \kev.
For a given source $m$ (defined by a radio-isotope together with one of its spatial component),
a given halo magnetic field configuration and a given escape fraction for SN Ia positrons, the storing of the final parameters allows us to calculate the steady state 511 \kev\ total annihilation flux:

\begin{equation}
\label{eq_flux}
F_{511}^{m}   =   \sum_{k=1}^{\mathrm{N_{0}}} \frac{ 2 \times (1-0.75f_{\mathrm{Ps},k})}{4 \pi d_{k}^{2}} \times \frac{\dot{N}_{\mathrm{e^{+}}}^{m}}{\mathrm{N_{0}}}~,  
\end{equation}
where $d_{k}$ and $f_{\mathrm{Ps},k}$ represent the distance of the annihilated positron $k$ to the Sun, and the total Ps fraction of the ISM phase in which positron $k$ annihilates \citep[calculated from][]{Guessoum05}. $\dot{N}_{\mathrm{e^{+}}}^{m}$ is the positron production rate for source $m$, reduced by the positron escape fraction.
 To obtain the total annihilation emission flux
due to all the nucleosynthesis positrons, we just need to sum over all sources $m$:
\begin{equation}
\label{eq_totalflux}
F_{511}   =   \sum_{m=1}^{\mathrm{M}} F_{511}^{m}~~~~,
\end{equation}
with M = 7, the number of possible sources ($m$ = \alu\ + CMZ, \alu\ + SFD, \titan\ + CMZ, \titan\ + SFD, \nickel\ + EB, \nickel\ + ED, \nickel\ + SFD).

%
\begin{table*}[!t]
\begin{center}
\caption{Bulge, disk and total 511 keV annihilation fluxes (in \funit) for \alu, \titan\ and \nickel\ positrons. 
\label{tab_flux} }
\scalebox{1}{\begin{tabular}{ccccrrrc}
\hline
\hline
\multirow{2}{*}{Source} & SN Ia escape  & Halo GMF & \centering{Bulge flux}  & Disk flux  & Total flux  & Bulge/Disk & Galactic escape   \\
 & fraction ($\%$) & configuration &  ($\times$10$^{-5}$) & ($\times$10$^{-4}$) &   ($\times$10$^{-4}$) &  ratio ($\%$) & fraction ($\%$)  \\
\hline
\multirow{3}{*}{$^{26}$Al} & \multirow{3}{*}{N/A} & N & 2.4 & 5.2 $\pm$  0.4 & 5.5 $\pm$  0.4 & 4.6 $\pm$  0.4 & 0.4 \\
& & D & 2.7 & 4.8 $\pm$  0.1 & 5.1 $\pm$  0.1 & 5.6 $\pm$  0.2 & 1.0 \\
& & X & 1.9 & 4.8 $\pm$  0.2 & 5.0 $\pm$  0.2 & 3.9 $\pm$  0.1 & 5.0 \\
\hline
\multirow{3}{*}{$^{44}$Ti} & \multirow{3}{*}{N/A} & N & 2.3 & 5.3 $\pm$  0.3 & 5.5 $\pm$  0.3 & 4.3 $\pm$  0.2 & 0.7 \\
& & D & 2.6 & 4.5 $\pm$  0.1 & 4.7 $\pm$  0.1 & 5.8 $\pm$  0.2 & 1.6 \\
& & X & 1.7 & 5.2 $\pm$  0.5 & 5.4 $\pm$  0.5 & 3.2 $\pm$  0.3 & 7.0 \\
\hline
\multirow{9}{*}{$^{56}$Ni} & \multirow{3}{*}{0.5$\%$} & N & 1.2 & 2.3 $\pm$  0.2 & 2.4 $\pm$  0.2 & 5.5 $\pm$  0.5 & 0.2 \\
& & D & 1.3 & 2.2 $\pm$  0.2 & 2.4 $\pm$  0.2 & 5.9 $\pm$  0.5 & 0.6 \\
& & X & 0.9 & 2.1 $\pm$  0.1 & 2.2 $\pm$  0.1 & 4.4 $\pm$  0.2 & 4.3 \\
\cline{2-8}
& \multirow{3}{*}{5$\%$} & N & 12.1 & 20.4 $\pm$  0.7 & 21.6 $\pm$  0.7 & 5.9 $\pm$  0.2 & 0.4 \\
& & D & 13.4 & 20.7 $\pm$  0.9 & 22.0 $\pm$  0.9 & 6.5 $\pm$  0.3 & 0.9 \\
& & X & 8.6 & 20.8 $\pm$  1.1 & 21.7 $\pm$  1.1 & 4.1 $\pm$  0.2 & 6.0 \\
\cline{2-8}
& \multirow{3}{*}{10$\%$} & N & 24.0  & 42.5 $\pm$  1.6 & 44.9 $\pm$  1.6 & 5.6 $\pm$  0.2 & 0.5 \\
& & D & 27.0 & 40.8 $\pm$  1.6 & 43.5 $\pm$  1.6 & 6.6 $\pm$  0.3 & 1.0 \\
& & X & 16.8 & 41.9 $\pm$  4.0 & 43.6 $\pm$  4.0 & 4.0 $\pm$  0.4 & 6.5 \\
\hline
\multirow{9}{*}{$^{56}$Ni+$^{44}$Ti+$^{26}$Al} & \multirow{3}{*}{0.5$\%$} & N & 5.9 & 12.8 $\pm$  0.6 & 13.3 $\pm$  0.6 & 4.7 $\pm$  0.2 & 0.5 \\
& & D & 6.6 & 11.5 $\pm$  0.3 & 12.2 $\pm$  0.3 & 5.8 $\pm$  0.1 & 1.2 \\
& & X & 4.5 & 12.2 $\pm$  0.6 & 12.6 $\pm$  0.6 & 3.7 $\pm$  0.2 & 5.7 \\
\cline{2-8}
& \multirow{3}{*}{5$\%$} & N & 16.8 & 30.9 $\pm$  0.9 & 32.6 $\pm$  0.9 & 5.5 $\pm$  0.2 & 0.5 \\
& & D & 18.7  & 30.0 $\pm$  0.9 & 31.8 $\pm$  0.9 & 6.3 $\pm$  0.2 & 1.0 \\
& & X & 12.2 & 30.9 $\pm$  1.3 & 32.2 $\pm$  1.3 & 3.9 $\pm$  0.2 & 6.0 \\
\cline{2-8}
& \multirow{3}{*}{10$\%$} & N & 28.7 & 53.0 $\pm$  1.7 & 55.9 $\pm$  1.7 & 5.5 $\pm$  0.2 & 0.5 \\
& & D & 32.3 & 50.1 $\pm$  1.6 & 53.3 $\pm$  1.6 & 6.5 $\pm$  0.2 & 1.0 \\
& & X & 20.3 & 52.0 $\pm$  4.1 & 54.0 $\pm$  4.1 & 3.9 $\pm$  0.3 & 6.4 \\
\hline
\end{tabular}}
\tablefoot{In the third column, N, D and X stand for no halo field, dipole and X-shape, respectively. For  $^{56}$Ni positrons, the fluxes are indicated for three different escape fractions from the SN Ia ejecta. The escape fractions indicated for the cumulated three radioactive sources apply only to $^{56}$Ni positrons. We also indicate the bulge-to-disk flux ratios and the fractions of positrons that escape the Galaxy. The uncertainties were calculated by a bootstrap method. The uncertainties are not shown for the bulge flux and the Galactic escape fraction because the maximum relative uncertainty is only $\sim$0.5$\%$ and $\sim$5$\%$, respectively.}
\end{center}
\end{table*}

\begin{figure*} [!ht]
\begin{center}
\begin{tabular}{cc}
\resizebox{9cm}{6.5cm}{\includegraphics{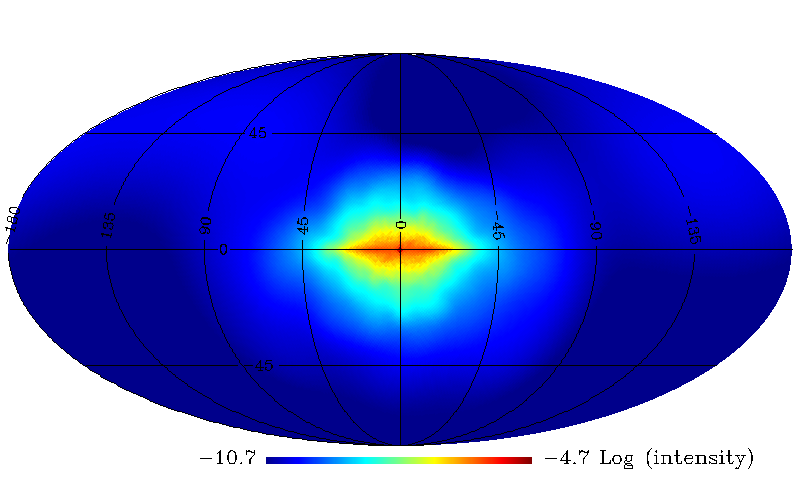}} & \resizebox{9cm}{6.5cm}{\includegraphics{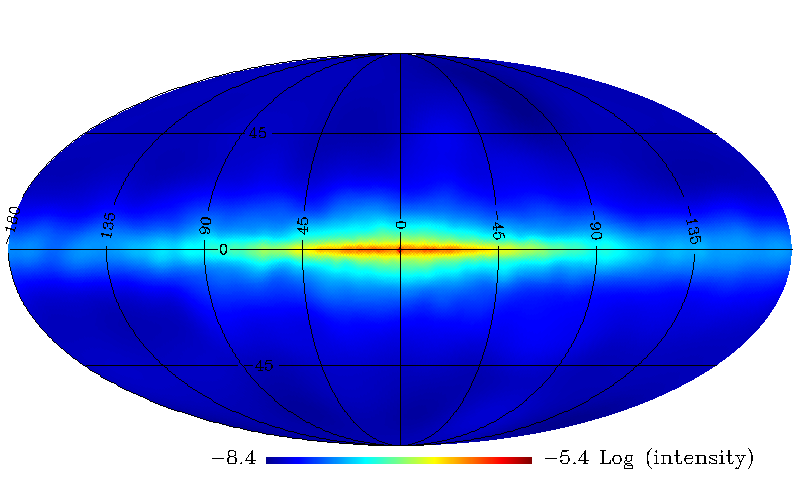}} \\
\resizebox{9cm}{6.5cm}{\includegraphics{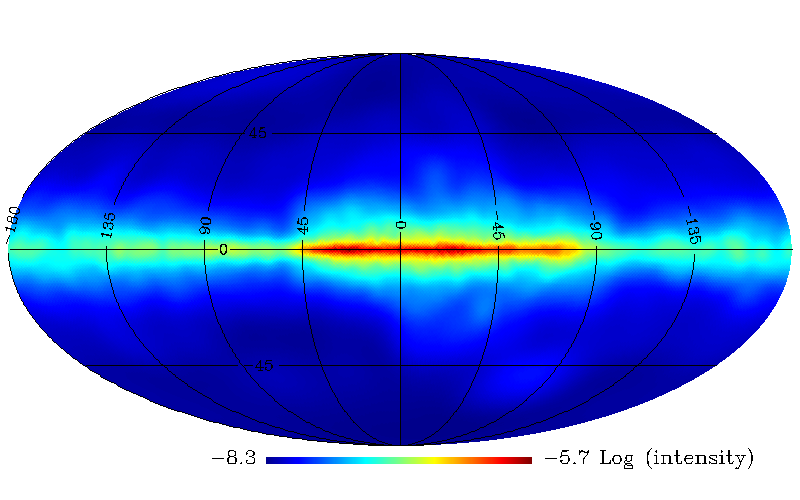}} & \resizebox{9cm}{6.5cm}{\includegraphics{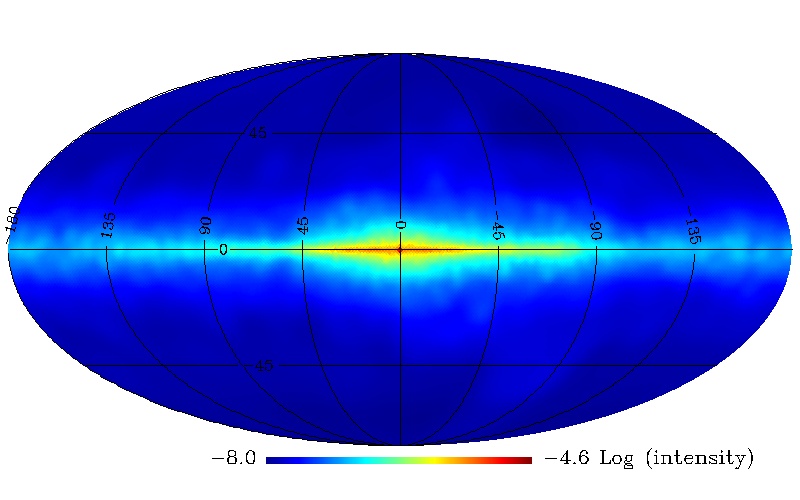}}
\end{tabular}
\caption{Simulated all-sky maps of the 511 \kev\ intensity distribution (in \iunit) for \nickel\ positrons
with the dipole halo field configuration and an SN~Ia escape fraction of 5\%. The maps correspond to the 511 \kev\ emission of positrons produced in the ellipsoidal bulge component (top left), the holed exponential component (top right), the SFD component (bottom left), and the entire Galaxy (by summing the three previous all-sky maps; bottom right).}
\label{fig_sn}
\end{center}
\end{figure*}


\section{Results of the simulations and discussion}
\label{simulations}

The numerical model described above allowed us to compute the annihilation emission associated with each radio-isotope and for each spatial component of its source distribution. The predicted annihilation emission could strongly depend on two poorly known parameters: the halo magnetic field configuration and the SN Ia escape fraction $f_{\rm{esc}}$ for \nickel\ positrons (see Sect. \ref{gmf} and \ref{sources}, respectively). 

In Sect. \ref{ism}, we introduced two representations for the distribution of phases in the ISM. All the simulations discussed below were performed for both representation and turned out to yield very similar results in terms of positron transport and morphology of the annihilation emission. Therefore, for these aspects, only the results corresponding to the random ISM model will be presented below. The only difference in the results obtained with the two prescriptions for the ISM lies in the annihilation phase fractions, and this will be discussed in Sect. \ref{annismphases}, where both sets of results will be shown.

We thus carried out a total of 39 simulations (=3$_{\mathrm{GMF}}$$\times$(2$_{\mathrm{^{26}Al}}$+2$_{\mathrm{^{44}Ti}}$
+  3$_{\mathrm{^{56}Ni}}$ $\times$3$_{f_{\rm{esc}}}$))  of 10$^{5}$ positrons
corresponding to all possible combinations of halo GMF configuration, source, and $f_{\rm{esc}}$. These simulations give 9 \mbox{(=$3_{\mathrm{GMF}}\times3_{f_{\rm{esc}}}$)} different total annihilation
emission sky maps due to all nucleosynthesis positrons. 
In Table \ref{tab_flux}, we present, for each positron source and each halo GMF configuration, the 511 \kev\ annihilation flux in the GB, the GD, and the entire Galaxy. The bulge-to-disk flux ratios and the fractions of positrons that escape from the Galaxy are also indicated.

In the following, we first expose the main results on the transport of positrons in each simulation, in particular their ranges and life times. 
Then, we present the predicted annihilation emission for each individual positron source and for all nucleosynthesis positrons together, depending on the GMF configuration. 
These models are then compared to recent measurements of the 511 \kev\ emission by the SPI spectrometer onboard the \mbox{INTEGRAL} mission, which shows that the observed emission cannot be completely accounted for. 
We therefore discuss in a subsequent part the possible contribution of a transient source at the GC to the 511\kev\ emission. Finally, we present the distribution of the positron annihilation over the different ISM phases and compare it to the spectrometric constraints.
 
\subsection{Positron ranges and life times}
\label{transport}

The numerical model allowed us to track the distance travelled by positrons 
in our Galaxy models. 
Two important results consistently emerged from our simulations.
First, for each positron source and for each halo GMF configuration, 
only a small fraction of positrons escape the Galaxy ($\lesssim 7\%$; 
see Table \ref{tab_flux}). 
In the case of an X-shape GMF, up to $30\%$ of the positrons produced by massive stars 
in the CMZ can escape the Galaxy, but these positrons represent only $10\%$ 
of the Galactic production by massive stars (see Sect. \ref{sources}). 
Second, whatever the positron source or the halo GMF configuration, positrons that do not escape the Galaxy only travel on average a distance \mbox{$\sim$1 kpc} from their injection site. 
 
 The travelled distance slightly varies with the halo GMF configuration and positron production sites. For instance, positrons produced in the very dense CMZ annihilate quickly and only travel about 150--300 pc, except in the simulation with the X-shape halo GMF where they travel on average 600 pc. This is because the vertical magnetic field lines of the X-shape halo GMF near the GC allow positrons to quickly escape the Galactic plane. The travelled distance also slightly varies with the initial energy of positrons. For instance, \nickel\ positrons from SNe Ia occurring in the SFD travel on average $\sim$600--700 pc, while the more energetic massive-star positrons (\alu\ and \titan\ positrons) produced in the SFD travel on average $\sim$0.9--1.1 kpc. But globally, nucleosynthesis positrons do not travel too far away from their birth places. This strongly explains: (a) the similar morphologies of the 511 \kev\ emission sky maps, as presented and discussed in Sect. \ref{comparison2data}, and (b) that our simulated 511 \kev\ spatial distributions closely reflect the spatial distributions of the positron sources.

%
\begin{figure} [!t]
\resizebox{\hsize}{6.5cm}{\includegraphics{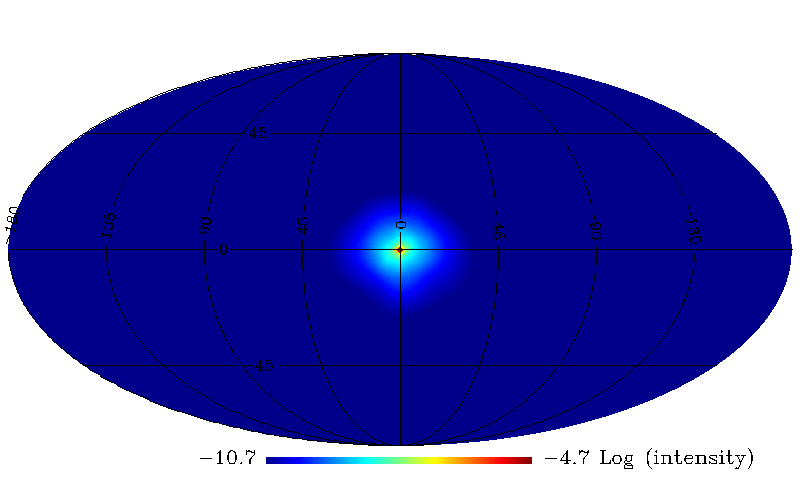}}
\resizebox{\hsize}{6.5cm}{\includegraphics{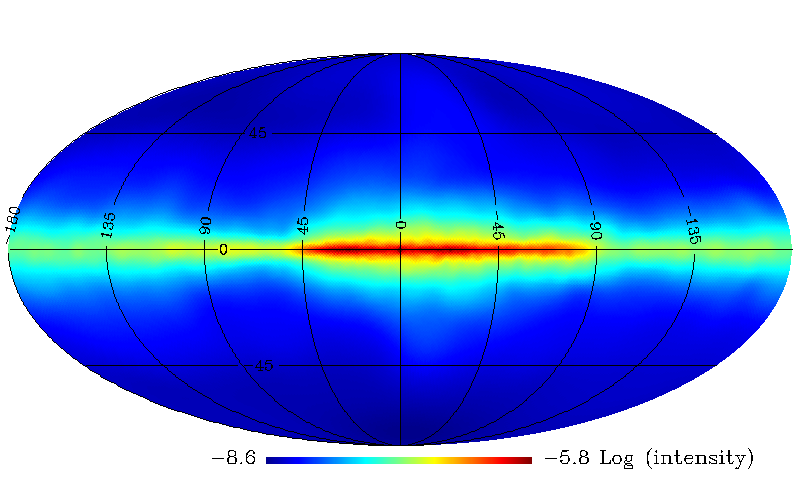}}
\resizebox{\hsize}{6.5cm}{\includegraphics{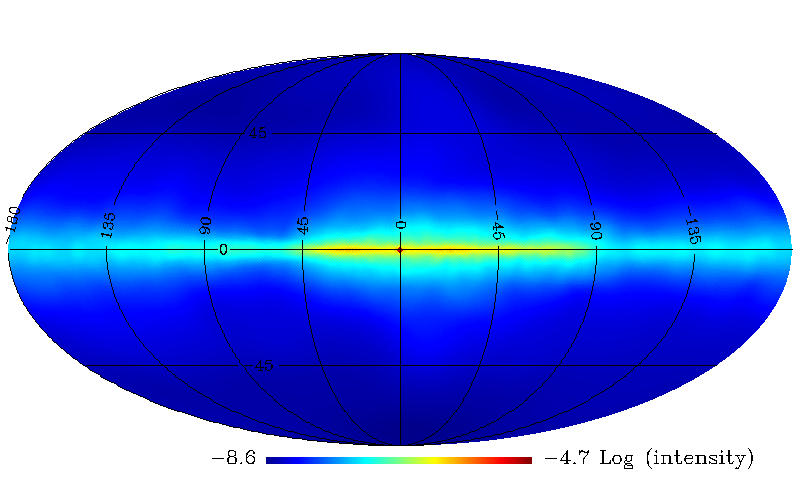}}
\caption{Simulated all-sky maps of the 511 \kev\ intensity distribution (in \iunit) for \alu\ and \titan\ positrons
with the dipole halo field configuration. From top to bottom, the maps correspond to the 511 \kev\ emission of 
positrons produced in the CMZ component, the SFD component, and in the entire Galaxy (by summing the two above all-sky maps).}
\label{fig_ms}
\end{figure}

%
\begin{figure} [!t]
\resizebox{\hsize}{6.5cm}{\includegraphics{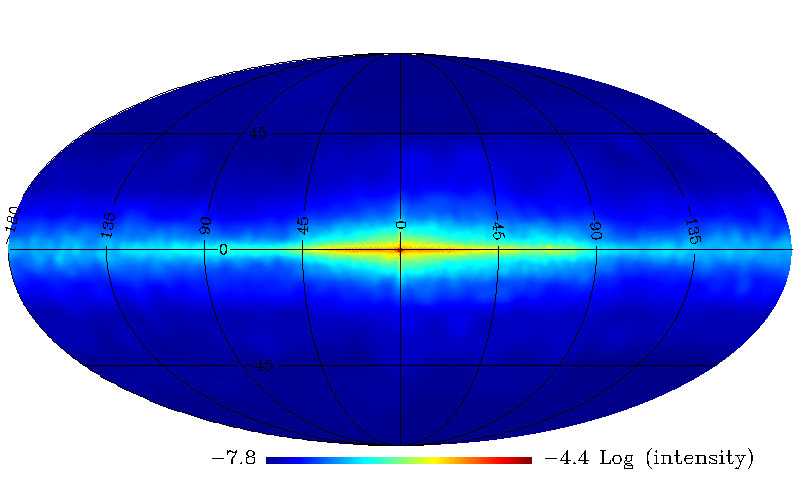}}
\resizebox{\hsize}{6.5cm}{\includegraphics{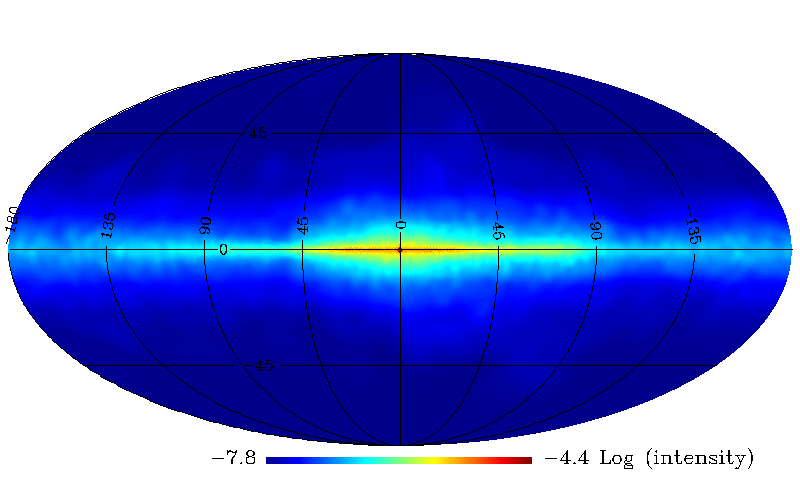}}
\resizebox{\hsize}{6.5cm}{\includegraphics{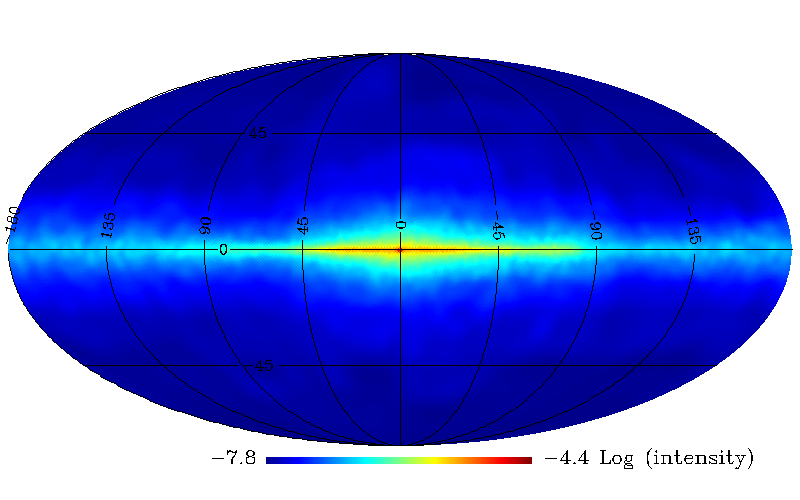}}
\caption{Simulated all-sky maps of the 511 \kev\ intensity distribution (in \iunit) for all nucleosynthesis positrons, 
with an SN Ia escape fraction of 5\% for \nickel\ positrons, and for the three halo GMF configurations. From top to bottom, the maps correspond to no halo field, the dipole field, and the X-shape field. The sky maps have been put on the same intensity logarithmic scale.}
\label{fig_allpos_dip_fesc5}
\end{figure}

The reasons why nucleosynthesis positrons do not travel far away from their birth places are: (a) the initial positron energies are a few 100 \kev\ only and not $\sim$1\mev\ as usually assigned in previous studies \citep[e.g.][]{Prantzos06} and (b) here, positrons are injected in a realistic ISM where the true densitiy of each ISM phase is taken into account, which also reduces the propagation distances. This is a stringent constraint. For instance, a positron entering the CNM near the Sun will travel in a medium with a true density $\simeq$40 cm$^{-3}$ contrary to the local CNM space-average density $\simeq$0.3 cm$^{-3}$ given by the model of \citet{Ferriere98}.

Accordingly, the average lifetime of nucleosynthesis positrons also slightly depends on the initial energy of positrons, the halo GMF configuration or the positron production sites. Whatever the halo field configuration and the escape fraction, the SN Ia positrons slow down in $\sim$(5--8)$\times$10$^{5}$ years when they are produced in the EB or the ED component, which consists mainly of tenuous Galactic regions (GB). In contrast, when they are produced in the denser regions of the SFD, they slow down in only $\sim$(2--4)$\times$10$^{5}$ years.
The more energetic massive-star positrons slow down on average in $\sim$(6--7)$\times$10$^{5}$ years when they are produced in the SFD. However, their mean lifetime depends on the halo GMF configuration when they are produced in the CMZ. With a dipole halo field, they slow down in only $\sim$1$\times$10$^{5}$ years, whereas they slow down on average 
in $\sim$7$\times$10$^{5}$ years with a X-shape halo field.

The simulations cannot explain the large bulge-to-disk (B/D) flux ratio of $\sim$1--3 derived from INTEGRAL/SPI observations \citep{Knodlseder05}. We only obtain B/D flux ratios of $\sim$0.05 for comparison. The reason is that the nucleosynthesis positrons produced in the GD cannot reach the GB.
Our simulations thus do not support the scenario proposed by \citet{Prantzos06}, who suggested that SN Ia positrons produced in the GD could be transported via a dipole GMF into the GB. However, the dipole halo field could have an important role confining positrons produced in the CMZ, as we will see in Sect \ref{comparison2data}.


\subsection{511\kev\ annihilation emission}
\label{skymaps}

Unless stated otherwise, all the results presented below were obtained for a dipole halo GMF and a SN Ia escape fraction $f_{\rm{esc}}$=5$\%$.

\begin{figure} [!t]
\centering
\resizebox{\hsize}{7.5cm}{\includegraphics{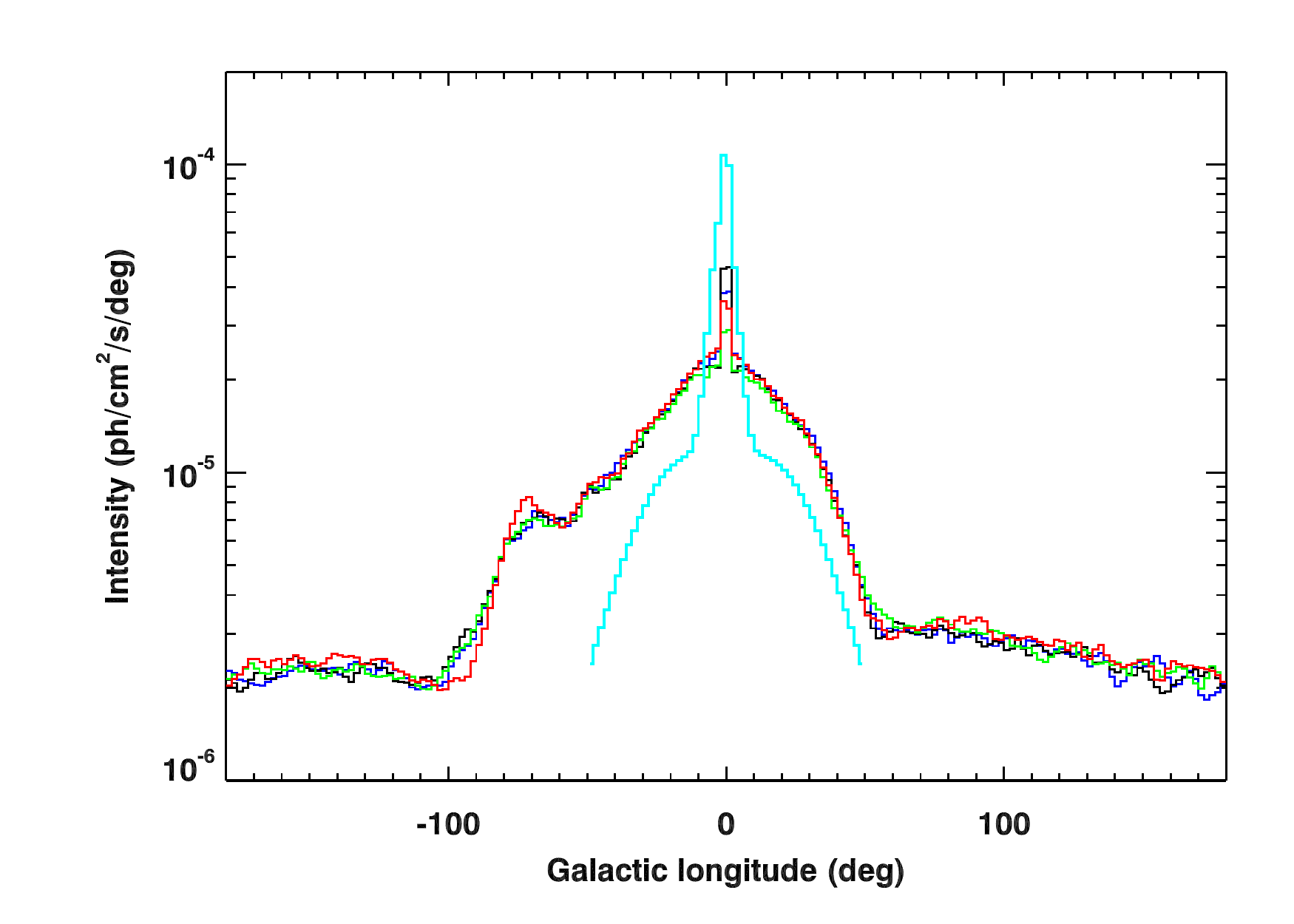}}
\caption{Longitude profiles of the 511 \kev\ emission of all nucleosynthesis positrons, with a SN~Ia 
escape fraction of 5\% for \nickel\ positrons, and for the three halo GMF configurations (corresponding to the maps in Fig. \ref{fig_allpos_dip_fesc5}, with an integration range $|b|\leq$~10$\degree$): no halo field (blue curve), dipole field (black curve), and X-shape field (green curve). The red curve corresponds to the longitude profile of all nucleosynthesis positrons in the absence of propagation, assuming they annihilate in a medium with a positronium fraction of 0.95.
The cyan curve corresponds to an analytical model obtained by model fitting to INTEGRAL/SPI observations, and given only over the longitude range where data are constraining (see text). The difference in normalisation shows that the positron injection rates used in the model are overestimated by a factor $\sim$2.}
\label{fig_longprof_allpos_dip_fesc5}
\end{figure}

Figures \ref{fig_sn} and \ref{fig_ms} show the all-sky intensity maps of the annihilation emission for \nickel\ positrons and \alu+\titan\ positrons, respectively (we present the cumulated emission from positrons from \alu\ and \titan\ because their respective contributions have the same morphology, see Fig. \ref{fig_longprof_allsrc_dip_fesc5}, because they have the same progenitors and similar injection energies). 
In these figures,  the 511 \kev\ intensity distribution of each spatial component of the source distribution is given before showing the total 511 \kev\ intensity distribution. 
Figure \ref{fig_allpos_dip_fesc5} shows the 511 \kev\ intensity distribution for the annihilation of all nucleosynthesis positrons for each halo GMF configuration and with $f_{\rm{esc}}$=5$\%$. The longitude profiles of these sky maps are presented in Fig. \ref{fig_longprof_allpos_dip_fesc5} while Fig. \ref{fig_longprof_allsrc_dip_fesc5} shows the longitude profile for each positron source and each $f_{\rm{esc}}$, in the case of a dipole halo GMF configuration. 

The halo GMF configuration has very little effect on the 511 \kev\ emission morphology, as illustrated by Fig. \ref{fig_allpos_dip_fesc5}. The longitude profiles (Fig. \ref{fig_longprof_allpos_dip_fesc5})  confirm this trend and underline that the 511 \kev\ emission spatial distribution reflects the positron source spatial distribution. We present in this figure the longitude profile of the 511 \kev\ emission corresponding to the case where positrons annihilate at their sources without propagation.
This profile is very similar to the three other profiles, which is explained by the fact that nucleosynthesis positrons do not propagate very far away from their birth places. 


For SN Ia positrons, the escape fraction has very little impact on the distribution of the resulting 511 \kev\ intensity sky maps because of slight differences in the energy spectra (see Sect. \ref{sources}). The main difference resides in 
the normalization of the 511 \kev\ intensity map, which depends on the positrons injection rate in the Galaxy 
(see Fig. \ref{fig_longprof_allsrc_dip_fesc5}). This can also be seen in Table \ref{tab_flux}, in which the total 511 \kev\ Galactic flux of each simulation is quasi-proportional to its 
$f_{\rm{esc}}$, for a given GMF configuration. 

In Fig. \ref{fig_allpos_dip_fesc5}, the bulk of the 511 \kev\ emission due to all nucleosynthesis positrons is concentrated in the longitude range $|l|\leq$~50$\degree$, in agreement with the extent measured by INTEGRAL/SPI \citep{Weidenspointner08a,Bouchet10}. The sky maps have a highly-peaked 511 \kev\ emission from the inner bulge, mostly due to the annihilation of \alu\ and \titan\ positrons produced in the CMZ. The main difference between the three GMF models comes from this emission component. 
The 511 \kev\ intensity in the inner bulge is $\sim$1.25 and  $\sim$2.5 times higher in the Galaxy model with a dipole halo GMF than in a Galaxy without a halo GMF and with an X-shape halo GMF, respectively. 

Another noteworthy feature concerning the morphology is the asymmetric emission from the GD. This is due to the annihilation emission of positrons produced in the SFD (see the same feature in Figs. \ref{fig_sn} and \ref{fig_ms}). The longitude extent of their GD emission is larger towards negative longitudes than toward positive longitudes. Most of the emission comes from between 0 to $l\simeq$ -75 $\degree$ at negative longitudes, whereas it comes from between 0 to $l\simeq$ 45 $\degree$ at positive longitudes (see also Figs. \ref{fig_longprof_allpos_dip_fesc5} and \ref{fig_longprof_allsrc_dip_fesc5}). This occurs because the line of sight towards $l\simeq$ -75$\degree$ follows over a long distance the positron-producing Sagittarius-Carina arm from the NE2001 model \citep[see Fig. 6 of][]{Cordes03}.

\subsection{Comparison to the INTEGRAL/SPI data}
\label{comparison2data}

We compared the simulated sky maps to INTEGRAL/SPI observations by a model fitting method, in which a sky model convolved by the instrument response function is fitted to the data together with a model of the instrumental background. The main results of a model fitting are the maximum likelihood ratio (MLR), which basically allows to compare different models, and the fit parameters, which in the present case correspond to the rescaling of the model required for a better match to the data \citep[see][fur further information on the model-fitting procedure]{Knodlseder05}. In this study, we used public data of the spectrometer SPI taken between December 9, 2002 and August 20, 2010. We performed the model fitting analysis in a 5 keV wide energy bin centered at 511 keV, including the Crab and Cygnus X-1 as point sources for the sake of completeness. We then compared our results to a phenomenological analytical model.

For the latter, we used the analytical model from \citet{Weidenspointner08b}. It is composed of a small spheroidal gaussian 
superimposed on a large spheroidal gaussian to account for the bulge emission (the inner and outer bulge, respectively), and a holed exponential disk as the one described by \citet{Robin03} for the young stellar population to represent the GD emission. We updated the spatial components (widths and positions of the gaussians, scale length and height of the disk) of their model from a fit to our data set. In contrast to the original model, we used a large gaussian slightly shifted toward negative longitudes \mbox{($\sim$ -1$\degree$)} because this improves the fit to the data \citep[see also][]{Skinner10}. In the following, this best-fit updated analytical model will be referred to as UW, and its  inner and outer spheroidal gaussian components as IB and OB, respectively. The longitude profile of this model is presented in Fig. \ref{fig_longprof_allpos_dip_fesc5}. It is plotted for the inner Galactic region only ($|l|\geq50\degree$) because it is currently unconstrained outside of this range \citep{Bouchet10}.

\begin{figure} [!t]
\centering
\resizebox{\hsize}{7.5cm}{\includegraphics{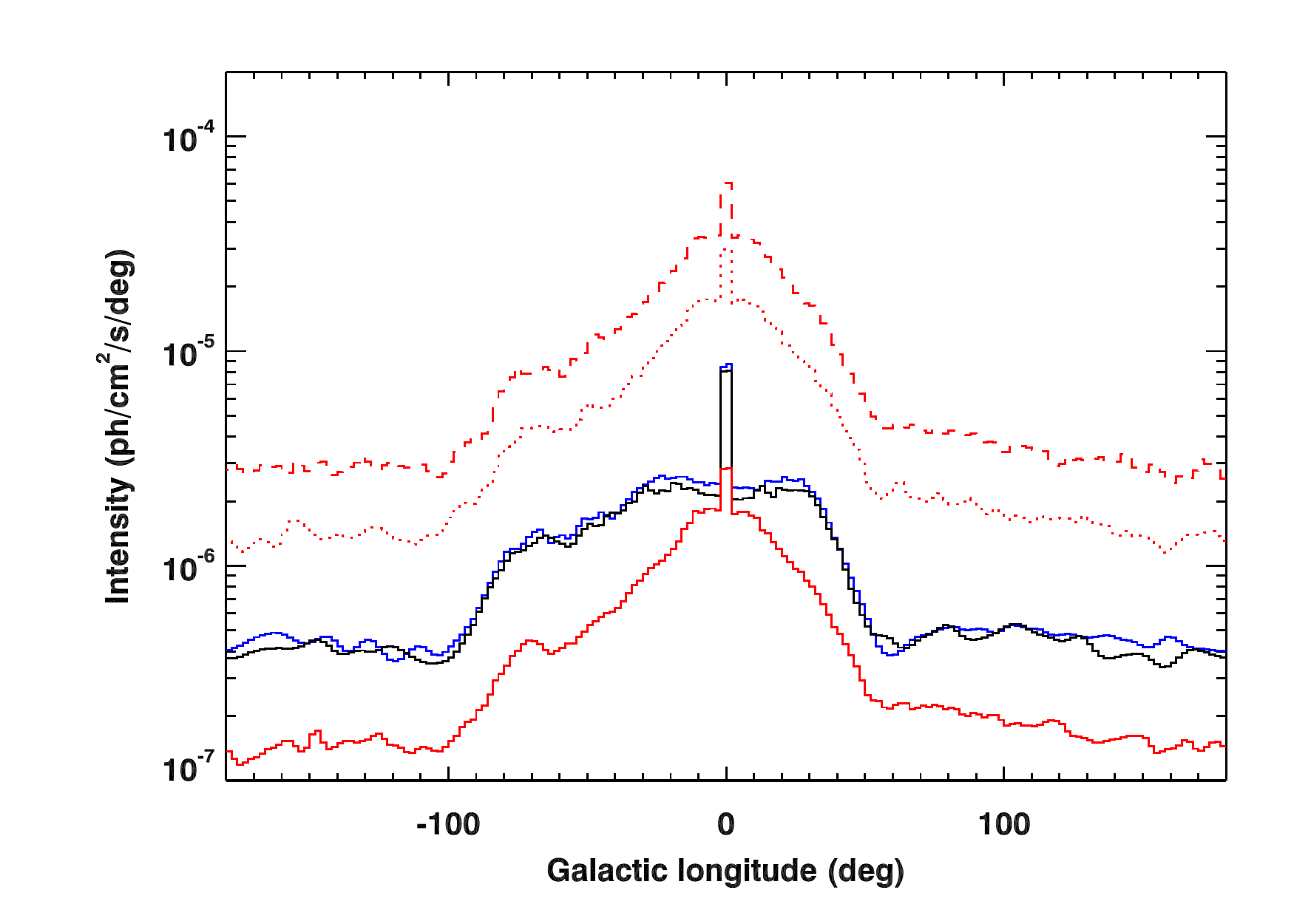}}
\caption{Longitude profiles of the 511 \kev\ emission (integration range $|b|\leq$~10$\degree$) from \alu\ (blue curve), \titan\ (black curve) and \nickel\ (red curves) positrons, with an SN~Ia escape fraction of 0.5\%\ (solid curve), 5\%\ (dotted curve) and 10\%\ (dashed curve) for \nickel\ positrons, in the case of a halo dipole field configuration.}
\label{fig_longprof_allsrc_dip_fesc5}
\end{figure}

Looking at Fig. \ref{fig_longprof_allpos_dip_fesc5}, it first appears that the normalisation of our models is too high. Reducing the positron injection rates used as a base case by a factor $\sim$2 would bring the predictions in line with the observations (a more quantitative discussion is given below). Yet, it is obvious that none of our intensity sky maps, whatever the halo GMF configuration, can fully account for the 511 \kev\ emission. Renormalizing the intensity would make it possible to account for the GD emission in the \mbox{$|l|=(10$-$50)\degree$} range, but in any case would not account for the full bulge emission. An additional source is thus necessary to explain the bulge emission, especially the outer bulge. However, in contrast to our previous study in \citet{Martin12}, we observe a strong and sharp intensity peak at the position of the GC region, which seems to match the inner bulge component. The reason is that we carefully modelled the CMZ and took into account the massive-star positrons produced there. Depending on the halo GMF configuration, the very dense CMZ is a major trap for (a) positrons that are directly injected in the CMZ and (b) positrons that are channeled into it (see the central intensity peak for each source in Fig. \ref{fig_longprof_allsrc_dip_fesc5}).


We confirmed by model fitting that our simulated 511\kev\ Galactic distributions can account for the observed GD emission as satisfactorily as the disk component of the best-fit analytical model. Fitting to the data the IB and OB components of the best-fit analytical model together with our simulated 511 \kev\ intensity distributions yields the results presented in Table \ref{tab_mlr}, for $f_{\rm{esc}}=5\%$ and the three halo GMF configurations (the fits for the other values of $f_{\rm{esc}}$ give similar MLR but with different scaling factors).
Statistically, our simulated emission models describe the measured disk emission as well as the young stellar population disk of the analytical best-fit model. The maximum difference in MLR is about 6, which is not significant. Therefore, the fit to the data does not allow us to constrain the halo GMF configuration. Moreover, whatever the halo field, the fitted fluxes of our simulated disk models are similar. The scaling factors for all our models are about 0.5. This means that the positron production rate of one or several radio-isotopes has perhaps been overestimated (see Sect. \ref{sources} for the uncertainties). 

The major difference between these different model fits lies in the fitted flux for the IB component, because our simulated 511 \kev\ intensity distributions already include a strong peak at the Galactic centre position, although at different levels depending the GMF. The IB component emission for a dipole halo field is  $\simeq$33$\%$ and $\simeq$40$\%$ lower than that for a no halo field and an X-shape halo field, respectively. The reason is that the dipole halo field confines positrons produced in the CMZ and makes them annihilate quickly, compared to the other two configurations (see Sect. \ref{transport}). Moreover, the dipole halo field is more likely to channel positrons produced by SNe Ia in the GB toward the CMZ.
Thus, with the dipole halo GMF, the inner bulge emission in our model is more intense and the analytical IB component superimposed on it needs to have a lower flux to bring the total emission up to the observed level.

In summary, our fits showed that the data cannot constrain the halo GMF configuration or the SN Ia escape fraction due to the large similarities between our different sky maps. 

We then sought to estimate independently the contribution of each positron source to the Galactic 511 \kev\ emission, and even beyond the contribution of some of their source components.To do this, we carried out a series of fits to the data with different emission models and/or combinations of models derived from the simulations with a dipole halo GMF. The results of the fits are shown in \mbox{Table \ref{tab_mlr_dipole}}. For massive-star positrons, we did not make any distinction between \alu\ and \titan\ because of their very similar intensity distributions. We thus used \alu\ models only and considered the more uncertain \titan\ as a possibility to increase the intensity by a factor of up to 2 (see Sect. \ref{sources}).

\begin{table}[!t]
\begin{center}
\caption{Results of the fits of our simulated 511 \kev\ emission spatial distributions due to all nucleosynthesis positrons, with a SN Ia escape fraction of 5$\%$ for \nickel\ positrons, and for the three halo GMF configurations, to about 8 years of INTEGRAL/SPI observations. \label{tab_mlr} }
\begin{tabular}{ccccc}
\hline
\hline
Model & Inner & Outer  & Disk & MLR \\
\hline
N & 1.07$\pm$0.17 & 5.6$\pm$0.3 & 17.1$\pm$1.3 & 2761.3  \\
D & 0.72$\pm$0.17 & 5.8$\pm$0.3 & 17.1$\pm$1.6 & 2760.9 \\
X & 1.18$\pm$0.17 & 5.7$\pm$0.3 & 17.9$\pm$1.6 & 2763.7 \\
UW            &  1.44$\pm$0.17 & 5.7$\pm$0.3 & 13.9$\pm$1.3 & 2766.6 \\
\hline
\end{tabular}
\tablefoot{The N, D and X models are those with no, dipole and X-shape halo fields, respectively.
The UW model is the updated best-fit analytical model. Columns 2-4 give the fitted flux in 10$^{-4}$ ph cm$^{-2}$ s$^{-1}$ for the inner bulge, outer bulge and disk components. Column 5 gives the maximum likelihood ratio. For the sake of completeness, the Crab and Cygnus X-1 fluxes in the 508.5--513.5\kev\ band were also fitted during the fit procedure.}
\end{center}
\end{table}

In a first step, we fitted to the data only one emission model due to positrons from one radio-isotope, all source components included. The fits are not good at all as illustrated by the longitude profiles of \mbox{Fig. \ref{fig_longprof_allsrc_dip_fesc5}}. In a second step, we added the analytical OB component to the nucleosynthesis positron all-sky models. Adding the OB component significantly improves the fits. The \mbox{511 \kev} all-sky emission from \alu\ positrons combined with the analytical OB component is able to explain the morphology of the observed all-sky 511 \kev\ emission as satisfactorily as the best-fit analytical model \mbox{(MLR=2758 and 2766, respectively)}. In contrast, the \nickel\ only emission model with the OB does not give such a good fit to the data (MLR=2743). This can be easily understood from Fig. \ref{fig_longprof_allsrc_dip_fesc5}: the intensity ratio between the emission peak
and the underlying disk is only $\sim$2 for \nickel\ positrons while it is $\sim$4 for massive-star positrons. This latter ratio is closer to the ratio of the best-fit analytical model which is $\sim$5. The \alu\ positron emission model and the OB can explain the morphology of the 511 \kev\ emission, but the \alu\ positron emission is rescaled in the fit by a factor of 3.6$\pm$0.3. Adding the contribution from \titan\ positrons to this model could not account for this scaling factor, even with the upper limits of the positron production rate by \alu\ and \titan. Therefore,\nickel\ positrons seem to be needed to explain quantitatively the observed 511 \kev\ emission in the disk and the inner bulge.


\begin{figure} [!t]
\centering
\resizebox{\hsize}{7.5cm}{\includegraphics{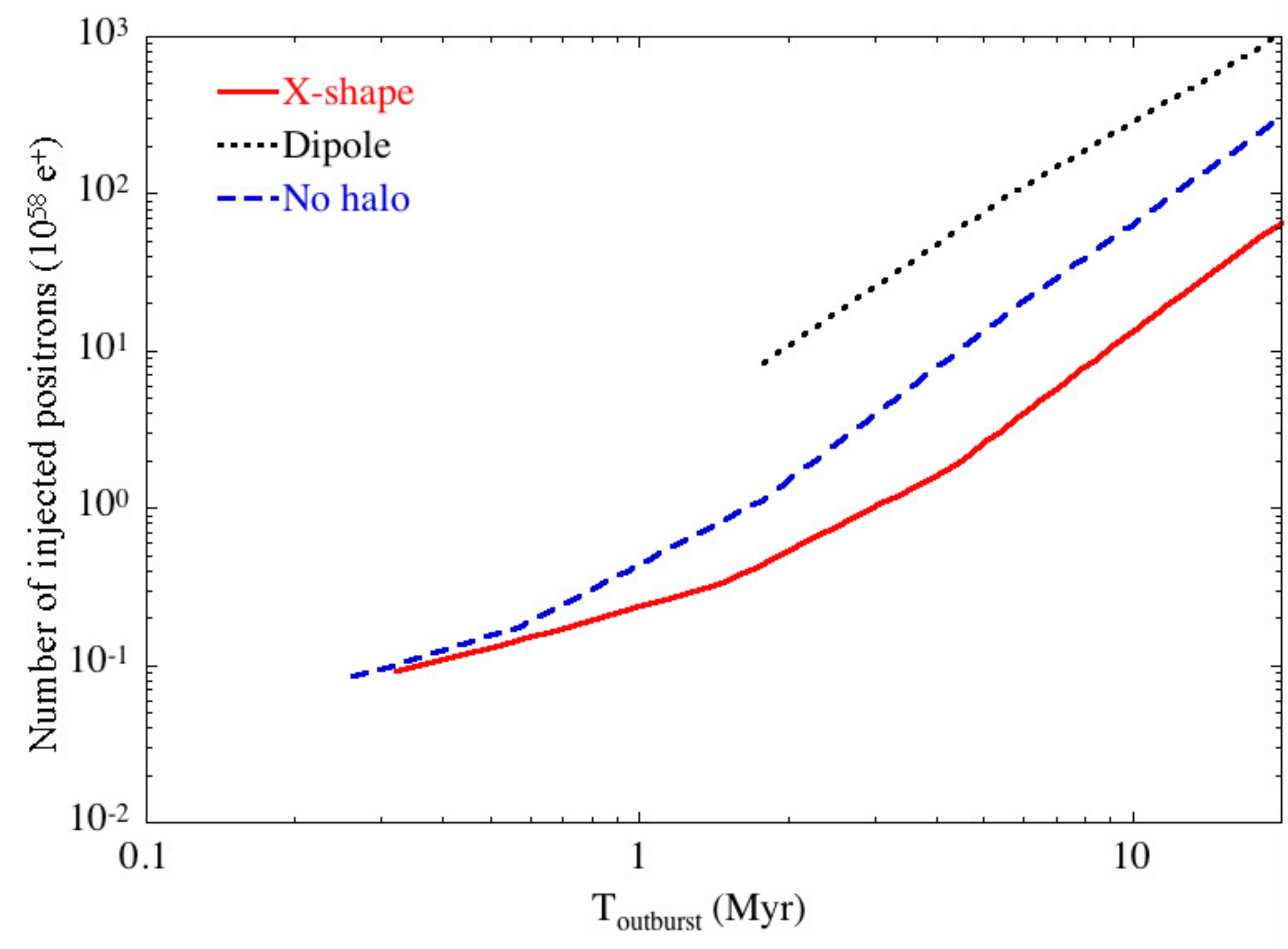}}
\caption{Number of positrons injected at the outburst in the CMZ, which makes it possible to reproduce a 511\kev\ flux of 5$\times 10^{-4}$\funit\ outside the CMZ (R>200 pc) today, as a function of time since outburst. The curves are shown only over the time intervals when the flux inside the CMZ (R<200 pc) is below the observed 1.5$\times 10^{-4}$ \funit\ (hence the truncations of the left parts). The solid red curve, dashed blue curve, and dotted black curve correspond to the simulations carried out with a X-shaped, no halo, and dipole halo GMF configurations, respectively.}
\label{fig_N-vs-T}
\end{figure}

To demonstrate this, in a last step, we fitted a model made of three components: (a) the emission of \alu\ positrons produced in the CMZ, (b) the emission of \alu\ positrons produced in the SFD added to the emission of all \nickel\ positrons, and (c) the OB component. This global model describes the data as satisfactorily as the UW model (MLR=2759.8). We obtain scaling factors of 3.4$\pm$0.6 and 0.62$\pm$0.06 for the CMZ \alu\ positron model and the (SFD \alu+\nickel) positron model, respectively.
The latter scaling factor suggests that one or several positron sources in the disk were overestimated. For instance, assuming that our estimate for the \alu\ positrons injection rate is correct, $f_{\rm{esc}}$$\simeq$2.5$\%$ would be sufficient to quantitatively explain the observed GD 511 from \alu\ and \nickel\ only. Taking into account a contribution from \titan\ would push the escape fraction even lower. Thus, it seems that the 511 \kev\ emission from the GD could be explained both morphologically and quantitatively from all radio-isotopes positrons. However, the 511\kev\ emission from the CMZ requires a larger correction factor of 3.4$\pm$0.6 to account for the inner bulge emission. 
One possibility would be that massive stars in the CMZ are either more numerous than assumed here (10$\%$ of all massive stars; see Sect. \ref{sources}) or more efficient at producing positrons, e.g., because of a more favorable IMF. However, we should keep in mind that the amount of \alu\ is constrained by the detection of the 1.8 \mev\ \gray\ line \citep[see e.g.][]{Martin09}. Another possibility would be to consider a contribution of SNe Ia in the CMZ. Assuming that 10$\%$ of the prompt SNe Ia occur in the CMZ as for massive stars (which roughly corresponds to the prompt SN Ia rate derived in the CMZ by \citealt{Schanne07} and \citealt{Higdon09}), we obtain a positron production rate of 
\begin{table*}[!th]
\begin{center}
\caption{Results of the fits of our simulated 511 \kev\ emission spatial distributions to about 8 years of INTEGRAL/SPI observations. \label{tab_mlr_dipole}}
\begin{tabular}{ccccc}
\hline
\hline
\multirow{2}{*}{Model} & Inner & Outer  & All-Sky & \multirow{2}{*}{MLR} \\
 & component & component  & component &  \\
\hline
\alu\ &  & & 45.3$\pm$1.0 (0.85) & 1987  \\
\nickel\ & & & 43.1$\pm$0.9 (1.93) & 2362 \\
All = (\alu+\titan+\nickel) &  &  & 44.8$\pm$1.0 (1.39)   & 2289 \\
\hline
OB+\alu\ &  & 7.1$\pm$0.3 & 18.3$\pm$1.4 (3.57) & 2758 \\
OB+\nickel\ & & 6.2$\pm$0.3 & 18.5$\pm$1.5 (0.82) & 2731 \\
OB+All~~    &  & 6.4$\pm$0.3 & 19.0$\pm$1.6 (0.59) & 2743 \\
\hline
CMZ \alu+OB+(SFD \alu+\nickel) & 0.8$\pm$0.1 (3.35) & 5.8$\pm$0.3 & 17.0$\pm$1.6 (0.62) & 2759.8 \\
\hline
UW (=IB+OB+Robin disk)            &  1.44$\pm$0.17 & 5.7$\pm$0.3 & 13.9$\pm$1.3 & 2766.6 \\
\hline
\end{tabular}
\tablefoot{Column 1 gives the model or the combination of models fitted to the data. The SN Ia escape fraction for \nickel\ positrons is 5$\%$ and the halo GMF is the dipole field. The \titan\ model was not fitted to the data because of the morphology of its emission is quite similar to that of \alu. Its contribution can be taken into account indirectly by scaling up the \alu\ contribution. Columns 2-4 give the fitted fluxes in 10$^{-4}$ ph cm$^{-2}$ s$^{-1}$ for the inner bulge, outer bulge and all-sky components. The numbers in parentheses are the scaling factors applied to our models in the fit. Column 5 gives the maximum likelihood ratio. The last line gives the results for the best-fit analytical model denoted UW, with IB and OB corresponding to the inner and outer gaussian-shaped bulge components.}
\end{center}
\end{table*}
$\simeq$6~$\times~10^{41}~\mathrm{e}^{+}/\mathrm{s}$. Added to the positron production rate of \alu\ positrons in the CMZ ($\simeq$3.2~$\times~10^{41}~\mathrm{e}^{+}/\mathrm{s}$), this contribution could explain the factor of $\sim$3 needed to account for the CMZ emission. Taking into account a contribution from \titan\ positrons in the CMZ would imply a lower escape fraction of SN Ia positrons of $f_{\rm{esc}}$$\leq$2.5$\%$, or a fraction of prompt SNe Ia occurring in the CMZ reduced by $\sim$2.

Finally, one should note that the above discussion holds for the dipole field model. The fit of the three-component model for the simulations with the other two halo GMF configurations gives similar MLRs (MLR=2760 and 2762.9 with no halo field and the X-shape halo field, respectively). However, in both cases, the emission model of \alu\ positrons produced in the CMZ needs a higher scaling factor than found with the dipole halo GMF (5.8 and 10 with no halo field and the X-shape halo field, respectively). The problem of the renormalisation of the \alu\ CMZ component thus becomes more acute, which opens the possibility of an additional contribution to the inner bulge emission on top of stellar nucleosynthesis.

In all cases, the positrons produced by steady state nucleosynthesis cannot explain the emission of the outer bulge detected by INTEGRAL/SPI.
One possibility recently investigated is that of dark matter scattering (\citealt{Vincent12}; note that this work did not include propagation which could modify the morphology of their 511\kev\ emission), which seems to confirm earlier predictions \citep[see e.g.][]{Boehm04}. In the following section, we present a possible alternative explanation based on a transient phenomenon.

As a last note on these considerations, and connected to what follows, we would like to emphasise that the separation of the observed bulge emission into an outer and an inner component is somewhat artificial and comes primarily from the choice of the functions used to model the observations \citep{Weidenspointner08b}. In the above paragraphs, we showed that the outer bulge emission cannot be reproduced and that the predicted inner bulge emission may also be short of what is measured. Actually, it may well be that a single component could account for both these two shortcomings at the same time (such as the transient component discussed below).

\subsection{Transient source}
\label{transient}

None of the positron radioactive sources, studied in a steady state way, can explain the $\simeq$10$\degree$ extended emission arising from the GB. In the following, we show that a 511\kev\ emission produced by a transient source injecting a large amount of positrons, seen at a particular moment after this event, hereafter called outburst, could explain the extended 511\kev\ emission from the GB. Only preliminary results are presented here, in order to illustrate the idea. A more detailed study with direct comparisons to the data via model fitting will be performed elsewhere.

We carried out simulations of the propagation of a given amount of \alu\ positrons produced in the CMZ during the outburst, for the three halo GMF configurations. We calculated 511\kev\ light curves from the inner bulge ($R$<200pc) and the outer bulge ($R$>200 pc), and compared them with the flux of the best-fit analytical model for two regions ($\simeq$1.5$\times 10^{-4}$ and $\simeq$5$\times 10^{-4}$ \funit\ in the inner bulge and outer bulge, respectively; see Table \ref{tab_mlr}). The 511 \kev\ flux  is computed taking into account the annihilation from Ps formed in flight and from thermalized positrons. Positrons that become thermalized can survive for several Myr in a tenuous plasma, and this is the case for the ionized ISM phases in the GB. The fractions of Ps formed in flight and the annihilation rates were taken from \citet{Guessoum05}. 

Figure \ref{fig_N-vs-T} shows the amount of positrons injected at the outburst that is necessary to obtain a flux $\simeq$5$\times 10^{-4}$ \funit\ outside the CMZ, as a function of time after the outburst. The curves are shown only over the time interval when the flux in the inner bulge is not greater than $\simeq$1.5$\times 10^{-4}$ \funit. 

%
\begin{figure} [!h]
\resizebox{\hsize}{6.5cm}{\includegraphics{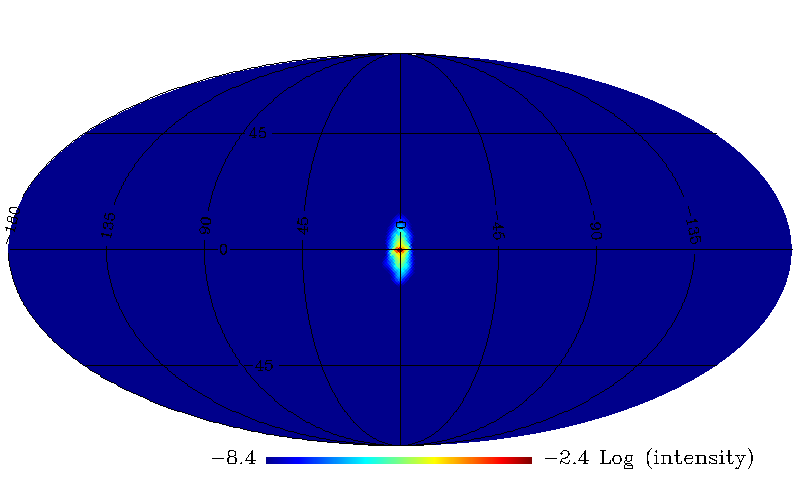}}
\resizebox{\hsize}{6.5cm}{\includegraphics{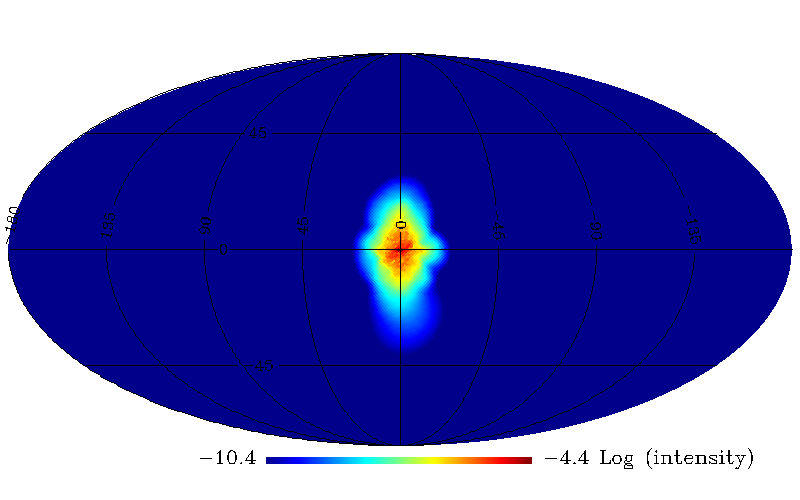}}
\resizebox{\hsize}{6.5cm}{\includegraphics{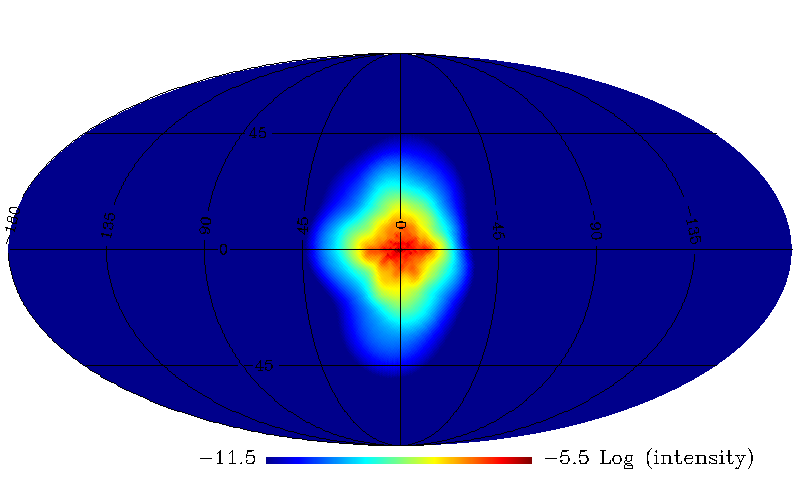}}
\caption{Simulated all-sky maps of the 511 \kev\ intensity distribution for an outburst of \alu\ positrons in the CMZ
for the X-shape halo field configuration (in \iunit). From top to bottom, the maps correspond to the average 511 \kev\ emission between 0 and 1$\times$10$^{5}$ yr, 5$\times$10$^{5}$ and 6$\times$10$^{5}$ yr, 1.9$\times$10$^{6}$ and 2$\times$10$^{6}$ yr after the outburst, respectively. The number of positrons injected in the outburst is 10$^{58}$. }
\label{fig_transient}
\end{figure}

We show that a very recent outburst which occurred less than 3$\times 10^{5}$ yr ago is ruled out because the 511 \kev\ emission remains too intense in the inner bulge compared to observations. However, whatever the halo GMF configuration, an outburst occurring between $\sim$2 and $\sim$10 Myr ago could explain the 511\kev\ flux outside the inner bulge. The number of positrons injected by this outburst ranges between 10$^{58}$ and 10$^{60}$ positrons. The X-shape halo GMF requires the lowest number of injected positrons to account for the outer bulge emission, whatever the time elapsed since the outburst. This is because the vertical magnetic field lines of the X-shape field allow positrons to escape quickly the CMZ, so that, the flux from the outer bulge increases more rapidly than that of the CMZ. Figure \ref{fig_transient} shows the temporal evolution of the morphology of the 511 \kev\ emission for an instantaneous injection of 10$^{58}$ \alu\ positrons in the CMZ, for the X-shaped halo GMF configuration. The 511 \kev\ emission is mainly concentrated in the CMZ region in the first 10$^{5}$ yr and then extends little by little into the outer GB.

The solution of an outburst that occurred > 0.3 Myr ago to explain the annihilation emission in the bulge matches with the starburst in the last $\sim$10 Myr in the CMZ that is suggested to be the origin of the large scale bipolar structures observed at several wavelengths \citep["Fermi bubbles" or/and WMAP Haze, see][]{Bland03,Su10,Law10,Carretti13}. It has also been suggested that these structures could have been generated by an outburst of the central supermassive black hole Sgr A*. However, based on observations of the radio lobes' morphology, \citet{Carretti13} concluded that they originate from a starburst event in the 200 pc diameter region around the GC rather than from the supermassive black hole. This does not rule out outbursts from Sgr A* as the source of the annihilation emission from the bulge, but this case involves other processes \citep[e.g.][]{Totani06,Cheng06,Cheng07} and positron energies for a release in a particular ISM. The fate of positrons ejected by Sgr A*, including their propagation, will be presented in another paper (Jean et al. 2014, in prep.).

\citet{Bland03} estimated that the energetics for the bipolar wind ($\sim$10$^{55}$ ergs) require a number of supernovae larger than 10$^4$. Assuming that a core-collapse supernova produces at least $\sim$10$^{52}$ positrons through the decay of $^{26}$Al and $^{44}$Ti, the number of supernovae needed $\sim$10 Myr ago to produce the measured 511 keV annihilation flux in the outer bulge is $\sim$10$^7$ for the X-shape halo GMF configuration (see Fig. \ref{fig_N-vs-T}). 
Such a number of supernovae is too large compared to that required by the energetics of the Fermi bubble. This suggests that an additional source of positrons is required. If hypernovae produce $\sim$10$^{55}$ positrons of $\sim$1 MeV per event, as suggested by \citet{Parizot05}, then a few thousand hypernovae that exploded 10 Myr ago can explain the measured flux in the outer bulge.


\subsection{Annihilation ISM phase fractions}
\label{annismphases}

%
\begin{table*}[!t]
\begin{center}
\caption{Annihilation phase fractions for $^{26}$Al, $^{44}$Ti, and $^{56}$Ni positrons, using the random ISM model.}
\label{tab_phase}
\scalebox{0.83}{\begin{tabular}{ccc@{\hspace{0.6cm}}ccccc@{\hspace{0.8cm}}ccccc@{\hspace{0.6cm}}c}
\hline
\hline
Source & SN Ia escape  & halo GMF & \multicolumn{5}{c}{Disk ISM phases}  &  \multicolumn{5}{c}{Bulge ISM phases}  & Sgr A* region   \\
 & fraction ($\%$) & configuration &  MM & CNM & WNM & WIM & HIM & MM & CNM & WNM & WIM & HIM &  IRH  \\
\hline
\multirow{3}{*}{$^{26}$Al} & \multirow{3}{*}{N/A} & N & 3.3 & 25.8 & 33.4 & 25.4 & 0.0 & 3.6 & 1.1 & 1.2 & 3.9 & 0.0 & 0.0 \\
& & D & 3.1 & 25.1 & 32.6 & 24.7 & 0.0 & 3.9 & 0.9 & 1.1 & 2.8 & 0.0 & 1.5 \\
& & X & 3.0 & 25.1 & 32.5 & 24.6 & 0.0 & 2.5 & 0.8 & 0.8 & 3.4 & 0.0 & 0.0 \\
\hline
\multirow{3}{*}{$^{44}$Ti} & \multirow{3}{*}{N/A} & N & 4.1 & 26.0 & 32.4 & 24.7 & 0.0 & 4.0 & 1.1 & 1.0 & 3.5 & 0.0 & 0.0 \\
& & D & 4.1 & 25.3 & 31.2 & 23.8 & 0.0 & 4.5 & 0.9 & 0.8 & 2.4 & 0.0 & 1.4 \\
& & X &3.9 & 25.1 & 31.2 & 23.4 & 0.0 & 2.7 & 0.8 & 0.6 & 2.6 & 0.0 & 0.0 \\
\hline
\multirow{3}{*}{$^{56}$Ni} & \multirow{3}{*}{5$\%$} & N & 1.8 & 21.3 & 34.1 & 29.3 & 0.0 & 0.8 & 0.6 & 0.7 & 8.4 & 0.0 & 0.0 \\
& & D & 1.9 & 20.7 & 33.6 & 28.2 & 0.0 & 1.8 & 0.7 & 1.2 & 7.4 & 0.0 & 0.8 \\
& & X & 1.6 & 20.2 & 32.7 & 27.4 & 0.0 & 0.6 & 0.4 & 0.4 & 6.1 & 0.0 & 0.0  \\
\hline
\multirow{3}{*}{$^{56}$Ni+$^{44}$Ti+$^{26}$Al}& \multirow{3}{*}{5$\%$} & N &2.4 & 22.6 & 33.7 & 28.1 & 0.0 & 1.7 & 0.8 & 0.8 & 7.0 & 0.0 & 0.0 \\
& & D & 2.4 & 22.1 & 33.1 & 27.0 & 0.0 & 2.5 & 0.8 & 1.1 & 6.0 & 0.0 & 1.0 \\
& & X & 2.2 & 21.7 & 32.5 & 26.4 & 0.0 & 1.2 & 0.5 & 0.5 & 5.2 & 0.0 & 0.0 \\
\cline{2-14}
\hline
\end{tabular}}
\tablefoot{In the third column, N, D and X stand for no halo field, dipole and X-shape field, respectively. The annihilation phase fractions are then given in \% for the Galactic disk, the Galactic bulge, and the Sgr A* region. For the latter region, only the annihilation fraction in the ionized radio halo (IRH) is given because the annihilation fractions in the other ISM components of the Sgr A* region are negligible. The sum was done over |z| $\leq$~1.5~kpc, so totals may slightly differ from 100\%.}
\end{center}
\end{table*}

Tables \ref{tab_phase} and \ref{tab_phase_onionskin}  present the distribution of positron annihilation over the various ISM phases. The results are shown for \nickel\ positrons with $f_{\rm{esc}}$=5$\%$, for all GMF models and for the two ISM prescriptions. The ISM phases in which positrons annihilate are little dependent on the halo GMF configuration and the SN Ia escape fraction. The ISM prescription, however, makes a difference. We start by presenting the results for the random ISM model before pointing out how the structured ISM model changes the picture.

Nucleosynthesis positrons annihilate mainly in the GD in the CNM, WNM and WIM, with fractions $\simeq$22$\%$, $\simeq$33$\%$ and $\simeq$27$\%$, respectively. 
As emphasised earlier, the majority of positrons produced in the GD do not propagate far away from their injection sites. The predominance of the warm phases can thus be explained from their relatively large filling factors, while the contribution of the CNM arise from the high densities of this ISM phase. Positrons do not annihilate in the HIM due to the very low density of this phase, confirming the estimates of \mbox{\citet{Jean06,Jean09}} and \citet{Churazov11}. 
 
Only $\sim$10$\%$ of all nucleosynthesis positrons annihilate in the GB. Within these $\sim$10$\%$, $\sim$66$\%$ annihilate in the WIM, $\sim$16$\%$ annihilate in the neutral atomic phases (CNM+WNM) and $\sim$18$\%$ in the MM. 

From spectrometric analyses, \citet{Churazov05} and \citet{Jean06} showed that the spectrum observed in the GB could be explained by annihilation predominantly in warm phases (one should note that these authors used different spatial morphologies for the modelling of the 511\,keV emission, so their estimates are not directly comparable). \citet{Churazov05} found that only WIM or a combination of CNM, WNM and WIM in similar proportions could explain all the emission. \citet{Jean06} found that WIM and WNM contributes both to the emission with a fraction of $\sim$50$\%$ each, without excluding a possible contribution of $\sim$20$\%$ from the CNM at the expense of the WNM. These estimates are roughly consistent with our predictions in terms of the order of importance of the phases: WIM, WNM, and CNM. The model sightly overpredicts the contribution of the MM phase (the situation is improved with the structured ISM model, see below). Yet, there are limitations to such a comparison: (1) as emphasised earlier, our simulations cannot reproduce the extended emission from the outer bulge while the spectrometric analyses mentioned above are based on the total inner and outer bulge signals; (2) the fractions derived by observations take into account some emission from the GD along the line of sight to the GB; (3) there are still serious uncertainties on the ISM filling factors in the GB \citep[see e.g.][]{Ferriere07}.

When using the structured ISM model, positron annihilation in the GD occurs mostly in the CNM, WNM and WIM, with fractions $\simeq$13$\%$, $\simeq$32$\%$ and $\simeq$37$\%$, respectively. The fraction of positrons that annihilate in the WNM remains similar compared to the simulations with the random ISM model, the fraction of positrons that annihilate in the CNM is reduced by $\sim$10$\%$, and the fraction in the WIM is increased by $\sim$10$\%$. This transfer can be understood from the prescription for the ISM layout: (1) in the structured ISM model, a positron injected into a spherical structure necessarily needs to go through a WIM before reaching a CNM, which lies deeper inside the spherical structure (see Sect. \ref{physics}); (2) in addition, due to the low mean filling factor of the CNM, the spherical shell of CNM is often quite thin, so that the positron does not stay there for a long time.

This global trend can also be observed for positrons annihilating in the GB. In this region, $\sim$77$\%$ of positrons annihilate in the WIM, $\sim$14$\%$ annihilate in the neutral atomic phases (CNM+WNM) and $\sim$9$\%$ in the MM. The fraction of positrons that annihilate in the neutral atomic phases remains roughly similar compared to the simulations with the random ISM model, the fraction of positrons that annihilate in the MM is reduced by $\sim$10$\%$, and the fraction in the WIM is increased by $\sim$10$\%$. 
With the structured ISM model in the CMZ, a positron escaping a HIM phase (the one with the largest filling factor) has no chance of entering a MM phase, which lies deep at the centre of the adopted spherical structure. In contrast, with the random ISM model, there was a non-negligible chance of moving from the HIM to the MM because of the $\sim$10\% filling factor of the latter. 


A more detailed study of the annihilation phase fractions will be done in the future by deriving 511\kev\ spectral distributions from our simulations, and comparing them to updated INTEGRAL/SPI observations.

%
\begin{table*}[!t]
\begin{center}
\caption{Same as Table \ref{tab_phase}, using the structured ISM model.}
\label{tab_phase_onionskin}
\scalebox{0.83}{\begin{tabular}{ccc@{\hspace{0.6cm}}ccccc@{\hspace{0.8cm}}ccccc@{\hspace{0.6cm}}c}
\hline
\hline
Source & SN Ia escape  & halo GMF & \multicolumn{5}{c}{Disk ISM phases}  &  \multicolumn{5}{c}{Bulge ISM phases}  & Sgr A* region   \\
 & fraction ($\%$) & configuration &  MM & CNM & WNM & WIM & HIM & MM & CNM & WNM & WIM & HIM &  IRH  \\
\hline
\multirow{3}{*}{$^{26}$Al} & \multirow{3}{*}{N/A} & N & 2.6 & 14.7 & 33.1 & 36.8 & 0.0 & 1.7 & 1.0 & 1.1 & 5.6 & 0.0 & 0.1 \\
& & D & 2.6 & 14.4 & 32.4 & 35.2 & 0.0 & 1.6 & 0.9 & 0.8 & 3.9 & 0.0 & 2.8 \\
& & X & 2.4 & 14.5 & 31.3 & 35.2 & 0.0 & 1.2 & 0.7 & 0.7 & 3.9 & 0.0 & 0.0 \\
\hline
\multirow{3}{*}{$^{44}$Ti} & \multirow{3}{*}{N/A} & N & 3.7 & 15.0 & 31.8 & 36.0 & 0.0 & 2.3 & 1.1 & 1.0 & 4.7 & 0.0 & 0.1 \\
& & D & 3.5 & 14.5 & 31.5 & 33.6 & 0.0 & 1.9 & 0.9 & 0.8 & 3.3 & 0.0 & 3.0 \\
& & X & 3.7 & 14.5 & 29.8 & 33.3 & 0.0 & 1.5 & 0.7 & 0.6 & 3.0 & 0.0 & 0.0 \\
\hline
\multirow{3}{*}{$^{56}$Ni} & \multirow{3}{*}{5$\%$} &N &1.3 & 11.2 & 31.7 & 42.4 & 0.0 & 0.4 & 0.5 & 0.6 & 8.1 & 0.0 & 0.0 \\
& & D & 1.2 & 10.8 & 31.2 & 40.8 & 0.0 & 0.6 & 0.6 & 0.8 & 7.3 & 0.0 & 1.9 \\
& & X & 1.4 & 10.9 & 30.0 & 39.3 & 0.0 & 0.3 & 0.3 & 0.4 & 5.5 & 0.0 & 0.0 \\
\hline
\multirow{3}{*}{$^{56}$Ni+$^{44}$Ti+$^{26}$Al} & \multirow{3}{*}{5$\%$} & N & 1.9 & 12.3 & 31.9 & 40.6 & 0.0 & 0.8 & 0.6 & 0.7 & 7.2 & 0.0 & 0.0 \\
& & D & 1.8 & 11.9 & 31.4 & 39.0 & 0.0 & 0.9 & 0.7 & 0.8 & 6.2 & 0.0 & 2.2 \\
& & X & 1.9 & 11.9 & 30.1 & 37.8 & 0.0 & 0.6 & 0.4 & 0.5 & 4.9 & 0.0 & 0.0 \\
\hline
\end{tabular}}
\end{center}
\end{table*}

\section{Summary and conclusions}
\label{conclusion}

The aim of this work was to determine if nucleosynthesis positrons could explain the morphology of the 511 \kev\ emission observed in our Galaxy.
Using a Monte Carlo code, we simulated their inhomogeneous collisional propagation and energy losses in a finely-structured ISM taking into account the Galactic magnetic field structure, the different ISM gaseous phases, particular features of the Galaxy such as the central molecular zone or the holed tilted disk, and testing two extreme ISM models.

We studied the contributions to the annihilation emission of positrons produced in massive stars (\alu\ and \titan) and in SNe Ia (\nickel). These sources have often been cited in the past as the most likely major contributors to Galactic positrons.
Due to large uncertainties in the positron escape fraction from the SNe Ia ejecta and the structure of the magnetic field in the Galactic halo, we tested several escape fractions and halo magnetic field configurations to see the impact of these parameters on the annihilation emission morphology.

The different combinations of these parameters experimented in the simulations result in quite similar 511 \kev\ emission morphologies. The main reason is that nucleosynthesis positrons do not propagate far away from their birth sites. In any case, a very low fraction of positrons $\lesssim$7$\%$ manage to escape the Galaxy. The rest of the positrons only travel on average $\sim$1 kpc and the 511 \kev\ intensity spatial distributions are thus strongly correlated with the source spatial distributions. Therefore, the steady state annihilation of nucleosynthesis positrons cannot account for the total annihilation emission observed in our Galaxy.

Comparison of our simulated sky maps to 8 years of INTEGRAL/SPI data confirms that nucleosynthesis positrons can explain the annihilation emission from the Galactic disk, but cannot fully account for that from the Galactic bulge. The morphology of the strongly peaked inner bulge emission could be explained by massive-stars positrons produced in the central molecular zone around the Galactic centre, with a possible contribution from positrons channelled there by a dipole halo field. However, depending on the magnetic field model, matching the observed intensity requires at least a contribution from prompt SNe Ia in the central molecular zone or more massive stars than currently, and at most the contribution from another unknown source. In any case, the emission from the outer bulge cannot be reproduced by steady state annihilation emission from nucleosynthesis positrons. We showed that a single and brief injection in the central molecular zone of a large amount of positrons, such as a starburst that occurred several Myr ago, could also explain the annihilation emission from the outer bulge. We found from our simulations that such an event occurring between 0.3 and 10 Myr ago and producing between 10$^{57}$ and 10$^{60}$ sub-MeV positrons could quantitatively explain the current emission from the outer bulge (and could also contribute to the inner bulge emission at some level). Nevertheless, detailed studies of these scenarios have to be undertaken before considering them as serious candidates to explain the complete 511 \kev\ annihilation emission from the Galactic bulge.

\begin{acknowledgements}
This paper is based on observations with INTEGRAL, an ESA project with instruments and science data centre funded by ESA member states (especially the PI countries: Denmark, France, Germany, Italy, Switzerland, Spain), Czech Republic and Poland, and with the participation of Russia and the USA. The SPI project has been completed under the responsibility and leadership of CNES/France.
Some of the results in this paper have been derived using the HEALPix \citep{Gorski05} package. 
\end{acknowledgements}

\bibliographystyle{aa}
\bibliography{article_aa}

\end{document}